\documentclass[11pt]{article}

\usepackage{bm} 
\usepackage[table]{xcolor}
\usepackage{authblk}
\usepackage{amsthm,amsmath,amsfonts,amssymb,mathtools} 
\usepackage[margin=1in]{geometry}
\usepackage{enumerate}
\usepackage{graphicx}
\usepackage{psfrag}
\usepackage{epsf}
\usepackage{hyperref}
\usepackage[utf8]{inputenc}
\usepackage[title]{appendix}
\usepackage[none]{hyphenat}
\usepackage{titlesec}
\usepackage{natbib}
\usepackage{color}
\usepackage{caption} 

\makeatletter
\providecommand{\leftsquigarrow}{%
  \mathrel{\mathpalette\reflect@squig\relax}%
}
\newcommand{\reflect@squig}[2]{%
  \reflectbox{$\m@th#1\rightsquigarrow$}%
}
\makeatother

\numberwithin{equation}{section}

\newtheorem{proposition}{Proposition}

\newtheorem{theorem}{Theorem}

\newtheorem{corollary}{Corollary}

\theoremstyle{plain}

\theoremstyle{definition}

\newcommand{\gauss}{\ensuremath{\textsc{gauss}}}
\newcommand{\trace}{\ensuremath{\text{\rm tr}}}

\newcommand{\hilde}[1]{{#1}^*}

\newcommand{\vm}{\ensuremath{\vec{\mu}}}
\newcommand{\vb}{\ensuremath{\vec{\beta}}}
\newcommand{\vg}{\ensuremath{\vec{\gamma}}}

\newcommand{\nullhyp}{\ensuremath{\cP}}
\newcommand{\althyp}{\ensuremath{\cQ}}
\newcommand{\nul}{\ensuremath{p}}
\newcommand{\alt}{\ensuremath{q}}

\newcommand{\cP}{\ensuremath{\mathcal{P}}}
\newcommand{\cQ}{\ensuremath{\mathcal{Q}}}

\newcommand{\commentout}[1]{}
\newcommand{\cH}{\ensuremath{\mathcal H}}
\newcommand{\cU}{\ensuremath{\mathcal U}}

\newcommand{\reals}{\ensuremath{{\mathbb R}}}

\newcommand{\meanspace}{\ensuremath{{\mathtt M}}}
\newcommand{\canspace}{\ensuremath{\mathtt{B}}}

\newcommand{\g}[1]{\ensuremath{#1}}
\newcommand{\gj}[2]{\ensuremath{#1}_{#2}}

\renewcommand{\vec}[1]{\ensuremath{{\bm #1}}}

\DeclareRobustCommand{\VANDER}[3]{#2}

\begin{document}
\author[12]{Peter Gr\"unwald}
\author[2]{Tyron Lardy}
\author[1]{Yunda Hao}
\author[3]{Shaul K. Bar-Lev}
\author[2]{Martijn de Jong}
\affil[1]{Centrum Wiskunde \& Informatica, Amsterdam, The Netherlands}
\affil[2]{Leiden University, Leiden, The Netherlands}
\affil[3]{Holon Institute of Technology, Holon, Israel}
\date{\today}
\title{Optimal E-Values for Exponential Families: the Simple Case}
\maketitle
\begin{abstract}
We provide a general condition under which e-variables in the form of a simple-vs.-simple likelihood ratio exist when the null hypothesis is a composite, multivariate exponential family. Such `simple' e-variables are easy to compute and expected-log-optimal with respect to any stopping time.
Simple e-variables were previously only known to exist in quite specific settings, but we offer a unifying theorem on their existence for testing exponential families.
We start with a simple alternative $Q$ and a regular exponential family null. Together these induce a second exponential family ${\cal Q}$ containing $Q$, with the same sufficient statistic as the null. 
Our theorem shows that simple e-variables exist whenever the covariance matrices of ${\cal Q}$ and the null are in a certain relation. 
A prime example in which this relation holds is testing whether a parameter in a linear regression is 0. Other examples include some $k$-sample tests, Gaussian location- and scale tests, and tests for more general classes of natural exponential families. While in all these examples, the implicit composite alternative is also an exponential family, in general this is not required. 
\end{abstract}

\section{Introduction}
\label{sec:introduction}
The work of Andrew Barron has been enormously influential in the development of {\em e-variables}, an alternative to the $p$-value that is suitable for designing hypothesis tests and confidence intervals with a flexible design, i.e. when sample sizes are not pre-specified, or when the decision to conduct new experiments may depend on past data, or even when the significance level $\alpha$ is set after observing the data \citep{Grunwald24}
--- \cite{ramdas2023savi} provides a recent overview of this exciting new area of statistical research, focusing on flexible design and {\em anytime-validity}, whereas \cite{ramdas2025hypothesistestingevalues} provides an extensive introduction focusing on the central building block, the e-variable, itself. A less technical introduction is provided by \cite{Ly25}. 
While Andrew himself has never published on e-variables, his seminal work on {\em Reverse Information Projection (RIPr) \/} \citep{LiB00} (see also the work by his Ph.D. students \cite{li1999estimation} and \cite{Brinda18}) is  the cornerstone of the math underlying \cite{GrunwaldHK23} (later extended by \cite{LardyHG24} and \cite{larsson2024numeraire}), which connects optimal e-variables to a RIPR onto the {\em null\/} hypothesis. 
On the other hand, there is Andrew's fundamental work on {\em universal coding and modeling\/ } (there are far too many papers to cite here ---  the earliest one may be \citep{ClarkeB90} while the most recent one is 
\citep{TakeuchiB24}). This work is directly connected to finding optimal e-variables for the {\em alternative\/} hypothesis, as explained by \cite{GrunwaldHK23,ramdas2023savi}. 
 Originally, much of it was done in the context of the Minimum Description Length {\em (MDL)\/} Principle \citep{BarronRY98}. For one of us (Gr\"unwald) MDL was the major research topic until around 2010 \citep{grunwald2007minimum}, while e-values have become his main topic since 2015 --- as such the influence of Barron's work on Gr\"unwald's work can hardly be overstated, and he would like to express his debt and gratitude.

In this paper we bring e-variables and in particular the RIPr together with another one of Andrew  Barron's central research interests: exponential families \citep{BarronS91,TakeuchiB97}. 
An important task is to test  whether a specific parameter in an exponential family is $0$ or not --- including linear regression testing as a special case.
Many classic tests are well-suited for this purpose~\citep{anderson1954test,lilliefors1967kolmogorov,stephens1974edf}. 
However, the vast majority of these methods are based on p-values, and thus designed for fixed sample size experiments. 
Here, we are instead interested in hypothesis tests that are based on e-values~\citep{GrunwaldHK23}, which is the value taken by an e-variable.
The most straightforward example of e-variables are likelihood ratios between simple alternatives and simple null hypotheses. 
E-variables for composite hypotheses, and in particular `good' e-variables, are generally more complicated.
However, e-variables in the form of a likelihood ratio with a single, special element of the null representing the full, composite null sometimes still exist.
We refer to such e-variables as `simple' e-variables. As we shall see below, their existence is intimately tied to properties of the aforementioned {\em RIPr}, connecting our work strongly to Barron's.

Simple e-variables, if they exist, can easily be computed, and are known to be optimal in an expected-log-optimality sense~\citep{KoolenG21,GrunwaldHK23}, the standard optimality measure for e-variables.
As such, it is desirable to find out whether or not simple e-variables exist in specific settings.
The main result of this paper, Theorem~\ref{thm:main}, provides a set of equivalent conditions under which simple e-variables exist for exponential family nulls.

In the remainder of this introduction, we provide a very short introduction to e-values (Section~\ref{sec:evars}) allowing us to to informally state our main result and provide an overview of the paper (Section~\ref{sec:overview}), followed by (Section~\ref{sec:formal_setting}) mathematical preliminaries, fixing notation, definitions and central concepts. 

\subsection{E-variables}\label{sec:evars}
Consider a family of distributions $\cP$ for data (random quantity) $U$. We think of $\cP$ as a null hypothesis, and we will use e-variables to gather evidence against $\nullhyp$.
An e-variable is a non-negative statistic with expected value bounded by one under the null, i.e. a non-negative statistic $S(U)$ such that $\mathbb{E}_{P}[S(U)]\leq 1$ for all $P\in \nullhyp$.
We give only a brief introduction to e-variables here and refer to e.g.~\citep{GrunwaldHK23,ramdas2023savi,ramdas2025hypothesistestingevalues} for detailed discussions, and to Appendix~\ref{app:happyreviewer} for a quick practical illustration. 
The realization of an e-variable on observed data will be referred to as an e-value, though the two terms are often used interchangeably. 
Large e-values give evidence against the null hypothesis, since by Markov's inequality we have that $P(S(U)\geq \frac1\alpha)\leq \alpha$ for any e-variable $S(U)$ and $P\in \nullhyp$. Thus, the test which rejects $\cP$ if $S(U) \geq \frac1\alpha$ has Type-I error probability bounded by $\alpha$, just as desired in classical testing. 
The focus here is on a static setting, where e-variables are computed for a single block of data (i.e. one observation of $U$).
However, the main application of e-variables is in anytime-valid settings, where data arrives sequentially and one wants a type-I error guarantee uniformly over time. 
Indeed, it is well-known that the product of sequentially computed e-variables again gives an e-variable, even if the definition of each subsequent e-variable depends on past e-values, which leads to an easy extension of the methods described here to such anytime-valid settings.
This is illustrated in Appendix~\ref{app:sequential} based on a simple $2$-sample testing scenario.

Since large e-values give evidence against the null, we look for e-variables that are, on average, `as large as possible' under the alternative hypothesis. 
In particular, we study growth-rate optimal (GRO) e-variables, an optimality criterion embraced implicitly or explicitly by most of the e-community \citep{ramdas2023savi};
\cite{GrunwaldHK23}, who introduced the concept, provide a simple comparison to standard statistical power. They define the GRO e-variable for single outcome $U$, relative to a simple alternative $\{Q\}$, to be the e-variable $S$ that, among all e-variables, 
maximizes the growth-rate  ${\mathbb E}_{U \sim Q}[\log S(U)]$ (also known as e-power \citep{wang2023ebacktesting,zhang2023existence}). 
In a celebrated result, \citeauthor{GrunwaldHK23} (see also \citep{LardyHG24,larsson2024numeraire}) show that, assuming $Q$ has density $q$ and every $P \in \cP$ has some density $p$, all relative to the same background measure, the GRO e-variable is given by:
\begin{equation}\label{eq:ripr}
    \frac{q(U)}{
p_{ \leftsquigarrow q}(U)},
\end{equation}
where $p_{ \leftsquigarrow q}$ denotes the density of the  reverse information projection of $Q$ on the convex hull of the null $\nullhyp$.
The reverse information projection of $Q$ on $\textsc{conv}(\nullhyp)$ is defined as the distribution that uniquely achieves $\inf_{P\in \textsc{conv}(\nullhyp)} D(Q\|P)$, which is known to exist whenever the latter is finite~\citep{li1999estimation,LardyHG24}.
Here, $D(Q\| P)$ denotes the Kullback-Leibler (KL) divergence between $Q$ and $P$, both defined as distributions for $U$.
In this article, all reverse information projection will be on $\textsc{conv}(\nullhyp)$, so we will not explicitly mention the domain of projection everywhere (i.e. referring to it simply as `the reverse information projection of $Q$').
The growth rate achieved by the GRO e-variable is given by
\begin{equation}
    \label{eq:GRO}
    \mathbb{E}_Q \left[ \log \frac{q(U)}{
p_{ \leftsquigarrow q}(U)}\right]= 
D(Q \| P_{\leftsquigarrow q} ) 
= \inf_{P \in \textsc{conv}(\nullhyp)} D(Q \| P).  
\end{equation}
In this paper we consider $\cP$ that are exponential families. Due to the fact that, with the exception of the Bernoulli and multinomial models, exponential families are not convex sets of distributions, finding the reverse information projection can be quite challenging \citep{Lardy21,HaoGLLA23}.
As our main result, we provide a simple and easily verifiable condition on $Q$ and exponential family $\cP$ under which 
\begin{equation}\label{eq:riprminb}
\inf_{P \in \textsc{conv}(\nullhyp)} D(Q \| P)  
= \min_{P \in\nullhyp}  D( Q\| P ),
\end{equation}
that is, the infimum is achieved by an element of $\nullhyp$, so that the problem greatly simplifies: the GRO e-variable is now the likelihood ratio $q(U)/p(U)$ between $Q$ and the $P \in \cP$ achieving the minimum. We will refer to such an e-variable as a {\em simple e-variable relative to $Q$}.
A big advantage of simple e-variables---besides their simplicity---is that their optimality extends beyond the static setting.
That is, suppose we were to observe independent copies $U_1,U_2,\dots$ of the data and assume that a simple e-variable of the form~\eqref{eq:simple} exists.
As with standard likelihood ratios, we can measure the total evidence as $\prod_{i=1}^n q(U_i)/p(U_i)$, which defines an e-variable for any fixed $n\in\mathbb{N}$. 
Instead of thinking of this as multiplication of individual e-variables, one can think of it as a likelihood ratio of $U_1,\dots,U_n$.
Proposition~\ref{prop:GRO} below then implies that $\prod_{i=1}^n q(U_i)/p(U_i)$ is the GRO e-variable for testing $\nullhyp$ against $\{Q\}$ based on $n$ data points. 
This statement shows that for any fixed sample size $n$, the best e-variable (in the GRO sense of~\ref{eq:GRO}) is the simple likelihood ratio.
Moreover, for applications where the sample size is not fixed beforehand,~\citet[Theorem 12]{KoolenG21} show that a more flexible statement is also true: if $\tau$ is any stopping time that is adapted to the data filtration, then $q(U^\tau)/p(U^\tau)$ is also a maximizer of $\mathbb{E}[\ln S_\tau]$ over all processes $S=(S_n)_{n\in \mathbb{N}}$ with $\mathbb{E}[S_\tau]\leq 1$.

%

\subsection{Main Result and Overview}
\label{sec:overview}

To give the reader a taste of what is to come, we now  briefly and still informally describe our main result Theorem~\ref{thm:main}, and we provide an overview of the rest of the paper.


We fix a regular multivariate exponential family 
%
$\cP$ for data $U$ with some sufficient statistic vector $X=t(U)$ and a distribution $Q$ for $U$, outside of $\cP$, and with density $q$. 
Here, $\cP$ will serve as the (composite) null hypothesis we are interested in testing, and $Q$ is the (simple) alternative hypothesis.
As our most important regularity 
conditions,
we assume (a) that $Q$ has a moment generating function and (b) that there
exists a distribution in $\cP$ that has the same mean for $X$, say $\vm^*$, as $Q$.
This distribution will be denoted by $P_{\vm^*}$ and its density by $p_{\vm^*}$. Assumption (a) is substantial:  essentially it means that tail probabilities are exponentially small, and thus, rules out, for example, situations in which outliers are a possibility. Assumption (b) is very weak, since our assumption that $\cP$ is a {\em regular\/} exponential family (as most exponential families encountered in practice are)  implies that for every $\vm^*$ in the interior of the convex hull of the support of $X$, $\cP$ contains a distribution $P_{\vm^*}$ with mean $\vm^*$ \citep{BarndorffNielsen78}. 
It is known that $P_{\vm^*}$ is the Reverse Information Projection (RIPr) of $Q$ onto $\cP$,  that is, it achieves $
\min_{P \in\nullhyp}  D( Q\| P )$.
It 
follows by Theorem~1 of \cite{GrunwaldHK23} that $q(U)/p_{\vm^*}(U)$ 
is an e-variable for testing $\cP$ if and only if (\ref{eq:riprminb}) holds, i.e. iff 
$\inf_{P \in \textsc{conv}(\nullhyp)} D(Q \| P)  
= \min_{P \in\nullhyp}  D( Q\| P )$.
Our theorem establishes a sufficient condition for when this is actually the case. 
It is based on constructing 
an auxiliary
exponential family ${\cal Q}$ with densities proportional to $\exp(\vb^T t(U)) q(U)$ for varying $\vb$: $\cQ$ contains $Q$ and has the same sufficient statistic as $\cP$. 
%
Letting $\Sigma_p(\vm)$ and $\Sigma_q(\vm)$ denote the covariance matrices of the $P_{\vm} \in \cP$ and $Q_{\vm} \in \cQ$ with mean $\vm$, Theorem~\ref{thm:main} below implies the following: under a further regularity condition on the parameter spaces of $\cP$ and $\cQ$, 
the likelihood ratio $q(U)/p_{\vm^*}(U)$ is an e-variable for testing $\cP$
whenever $\Sigma_p(\vm) - \Sigma_q(\vm)$ is positive semidefinite for all $\vm$ in the mean-value parameter space of $\cQ$ (additionally, three equivalent conditions will be given). 
If this happens, then we may further conclude that for  {\em every\/} element $Q_{\vm'}$ of the constructed $\cQ$, the likelihood ratio $q_{\vm'}(U)/p_{\vm'}(U)$ is an e-variable.
%

\begin{figure}[t]
    \centering
    \includegraphics[width=0.5\linewidth]{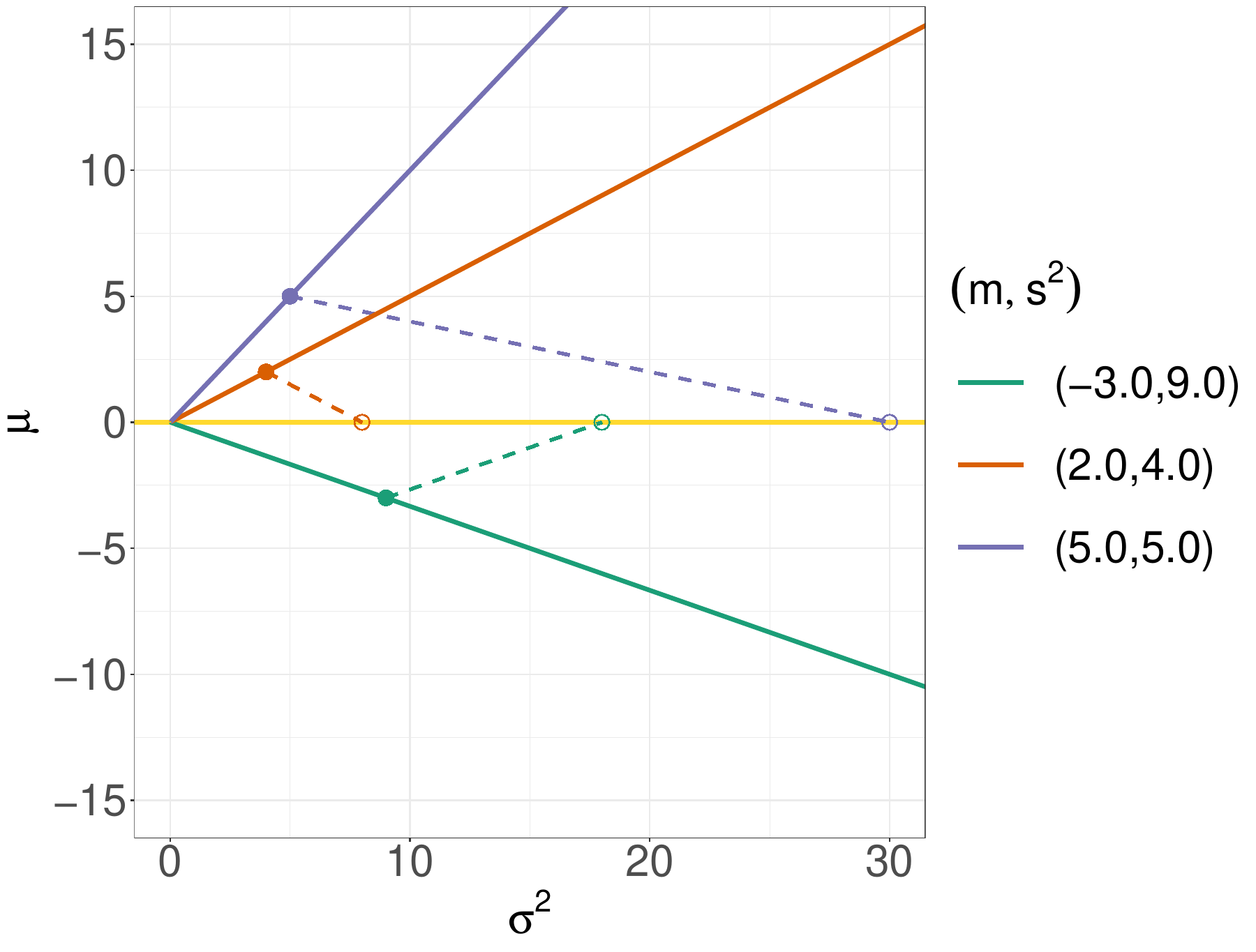}
    \caption{The family $\althyp$ for various $(m,s^2)$. The coordinate grid represents the parameters of the full Gaussian family, the horizontal line shows the parameter space of $\nullhyp$, the sloped lines show the parameters of the distributions in $\althyp$, and the dashed lines show the projection of $(m,s^2)$ onto the parameter space of $\nullhyp$. For example, we may start out with $Q$ expressing $U \sim N(m,s^2)$ with $m= -3.0, s^2 =9.0$, represented as the green dot on the green line. Its RIPr onto $\cP$ is the green point on the yellow line. The corresponding family $\cQ$, constructed in terms of $Q$ and $\cP$, is depicted by the green solid line. The theorem implies that the likelihood ratio between any point on the green line and its RIPr onto the yellow line is an e-variable; similarly for the red and blue lines.}
    \label{fig:gausscaleQfam}
\end{figure}

For example, suppose that, under $Q$, $U \sim N(m,s^2)$ for fixed $m, s^2$ and $\cP = \{N(0,\sigma^2): \sigma^2 > 0 \}$ is the univariate (scale) family of normal distributions. Then $X=U^2$ is the sufficient statistic for $\cP$. $X$  has mean $\vm^*=s^2 + m^2$ under $Q$,
and now Theorem~\ref{thm:main} can be used to show that $p_{m,s^2}(U)/p_{0, s^2+m^2}(U)$ is an e-variable for testing $\cP$. 
Here, $p_{m,s^2}$ denotes the density of the normal distribution with mean $m$ and variance $s^2$.
Moreover, the auxiliary family $\cQ$ is given by a subset of the full Gaussian family, and each element of $\cQ$ gives rise to a different e-variable for testing $\cP$.
This situation is illustrated in Figure~\ref{fig:gausscaleQfam} and is treated in detail in Section~\ref{sec:gausscale}, and extended to linear regression testing -- arguably our most important application -- in Section~\ref{sec:regression}.

We stress that, while our approach starts with a simple alternative $Q$, the results are still applicable if one is interested in a composite alternative $\cH_1$.
To this end, take any $Q \in \cH_1$ and use our main result to determine whether a simple e-variable with respect to $Q$ exists. 
If one exists for every $Q\in \cH_1$, an e-variable for the full alternative can easily be constructed either by the method of mixtures (see Appendix~\ref{app:happyreviewer}) or the prequential (sequential plug-in learning) method \citep{ramdas2023savi}.
In some cases, but not all, the auxiliary exponential family $\cQ$ constructed for any $Q\in \cH_1$ is equal to $\cH_1$ (Appendix~\ref{app:happyreviewer} explicitly considers  a fixed null $\cP$ with two different practically relevant $\cH_1$, one equal to $\cQ$ and one different).
In the case that $\cQ=\cH_1$, verifying whether a simple e-variable with respect to any $Q$ exists using Theorem~\ref{thm:main} is equivalent to checking whether a simple e-variable with respect to every element of $\cH_1$ exists.
%
Another simplifying situation is when $\cH_1$
can be 
partitioned as $\cH_1 = \bigcup_{\theta \in \Theta} \cQ^{(\theta)}$
in such a way that for each $Q \in \cH_1$, the associated family $\cQ$ constructed from $\cP$ and $Q$ is equal to $\cQ^{(\theta)}$ for some $\theta$. 
To apply the method of mixtures, it is not necessary to check whether an e-variable exists for every $Q\in \cH_1$, but simply to check the conditions for each $\cQ^{(\theta)}$.
As is suggested by   Figure~\ref{fig:gausscaleQfam}, 
this  happens, for example, in the Gaussian scale example of 
Section~\ref{sec:gausscale}, if we consider as alternative $\cH_1$ the full Gaussian family. 
We can start with any $Q= N(m,s^2)$ and generate $\cQ$ which then coincides with some $\cQ^{(\theta)}$, corresponding to a specific sloped line in the figure.
Together, all these sloped lines span $\cH_1$.
In fact, it turns out that a natural choice of $\cH_1$ that partitions into $\cQ^{(\theta)}$ is possible in {\em all\/} our examples, and that this $\cH_1$ is itself an exponential family in all these examples. Nevertheless, we stress that in general our method does not in any way require $\cH_1$ to be an exponential family --- only $\cP$ is required to be so. 

 
In the remainder of this introductory section, we fix notation and definitions of exponential families and e-variables. 
In Section~\ref{sec:simpleH1general} we show how, based on the constructed family $\cQ$, one can often easily construct {\em local\/} e-variables, i.e. e-variables with the null restricted to a subset of $\cP$. Then, in Section~\ref{sec:main} we present our main theorem, extending the insight to global e-variables. Section~\ref{sec:examples} provides several examples. This includes cases for which simple e-variables were already established, such as certain k-sample tests~\citep{TurnerLG24,HaoGLLA23}, multivariate Gaussian location with fixed covariance matrix \citep{SpectorCL23} or --- in an unpublished master's thesis --- the linear regression model \citep{DeJong21}, as well as cases for which it was previously unknown whether simple e-variables exist, such as for a broad class of natural exponential families. Theorem~\ref{thm:main} can thus be seen as a unification and generalization of known results on the existence of simple e-variables, leading to deeper understanding of why they sometimes exist.
Section~\ref{sec:simpleH1generalproofs} provides the proof for Theorem~\ref{thm:main}, which is based on  convex duality properties of exponential families. 
Finally, Section~\ref{sec:conclusion} provides a concluding discussion and points out potential future directions.  

\subsection{Formal Setting}\label{sec:formal_setting}
We study general hypothesis testing problems in which the null hypothesis $\nullhyp$ is a regular (and hence full) $d$-dimensional exponential family. 
Here and in the sequel, we will freely use standard properties of exponential families without explicitly referring to their definitions and proofs, for which we refer to e.g. \citep{BarndorffNielsen78,brown1986fundamentals,efron_2022}.
Each member of $\nullhyp$ is a distribution for a random element $\g{U}$, that takes values in some set $\cU$, with a density relative to some given underlying measure $\nu$ on $\cU$.
The sufficient statistic vector is denoted by $\g{X} = (\gj{X}{1}, \ldots, \gj{X}{d})$ with $\gj{X}{j}= t_j(\g{U})$ for given measurable functions $t_1, \ldots, t_{d}$.
We furthermore define $\meanspace_{\nul}$ to be the mean-value parameter space of $\nullhyp$, i.e. the set of all $\vm$ such that $\mathbb{E}_P[\g{X}] = \vm$ for some $P \in \nullhyp$.
For any $\vm \in \meanspace_{\nul}$, we denote by $P_{\vm}$  the unique element of $\nullhyp$ with $\mathbb{E}_{P_{\vm}}[X]=\vm$, so that $\nullhyp=\{P_\vm: \vm\in \meanspace_{\nul}\}$.
As usual, this parameterization of $\nullhyp$ is referred to as its mean-value parameterization. 
Furthermore, we use $\Sigma_\nul$ to denote the variance function of $\nullhyp$. 
That is, for all $\vm \in \meanspace_\nul$, $\Sigma_{\nul}(\vm)$ is the covariance matrix corresponding to $P_{\vm}$.

Since $\nullhyp$ is an exponential family, the density of any member of $\nullhyp$ can be written, for each fixed $\vm^*\in \meanspace_{\nul}$, as  
\begin{equation}\label{eq:exp_fam_dens}
p_{\vb; \vm^*}(u)
= \frac{1}{Z_{\nul}(\vb; \vm^*)} \exp \left(\sum_{j=1}^d \beta_j t_j(u) \right) \cdot p_{\vm^*}(u),
\end{equation}
where $Z(\vb; \vm^*) = \int \exp(\sum \beta_j t_j(u)) p_{\vm^*}(u) d \nu$, and $\vb \in \reals^d$ such that $Z_\nul(\vb;\vm^*)<\infty$. 
Therefore, $\nullhyp$ can also be parameterized as  $\nullhyp = \{P_{\vb; \vm^*}: \vb \in \canspace_{\nul, \vm^*}\}$, where $\canspace_{\nul; \vm^*} \subset \reals^d$ denotes the canonical parameter space with respect to $\vm^*$, i.e. the set of all $\vb$ for which $Z_\nul(\vb;\vm^*)<\infty$.
We use $\vb_\nul(\vm'; \vm^*)$ to denote the $\vb\in \canspace_{\nul, \vm^*}$ such that $\mathbf{E}_{P_{\vb;\vm^*}}[X]=\vm'$ and set $\vm_\nul(\cdot; \vm^*) = \vb^{-1}_\nul(\cdot; \vm^*)$ to be its inverse. $\vb_\nul(\cdot ; \vm^*)$ maps mean-value parameters to corresponding canonical parameters and $\vm_\nul(\cdot; \vm^*)$ vice versa.
Note that $p_{\vm^*} = p_{{\bf 0}, \vm^*}$, and that we can see from the notation (one versus two subscripts) whether a density is given in the mean- or canonical representation, respectively. 

The reason for explicitly denoting the mean $\vm^*$ of the carrier density, which is unconventional, is that it will be convenient to simultaneously work with different canonical parameterizations, i.e. with respect to a different element of $\meanspace_\nul$, below. 
These are all linearly related to one another in the sense that for each $\vm_1, \vm_2\in \meanspace_\nul$, there is a fixed vector $\vec{\gamma}$ such that for all $\vb \in \canspace_{p, \vm_1}$ it holds that $p_{\vb; \vm_1} = p_{\vb + \vec{\gamma}; \vm_2}$. 
This can be seen by taking $\vec{\gamma}=-\vb_\nul(\vm_2;\vm_1)$, since one then has
\begin{equation}\label{eq:lin_rela}
\begin{split}
    p_{\vb;\vm_1}(u) 
    &=\frac{1}{Z_p(\vb;\vm_1)}\exp\left(\sum_{j=1}^d (\beta_j+\gamma_j) t_j(u) \right)\exp\left(\sum_{j=1}^d  -\gamma_j t_j(u) \right) p_{\vm_1}(u)\\
    &= \frac{Z_p(-\vec{\gamma};\vm_1)}{Z_p(\vb;\vm_1)}\exp\left(\sum_{j=1}^d (\beta_j+\gamma_j) t_j(u) \right)p_{-\vec{\gamma};\vm_1}(u)\\
    &=\frac{1}{Z_p(\vb+\vec{\gamma};\vm_2)} \exp\left(\sum_{j=1}^d (\beta_j+\gamma_j) t_j(u) \right)p_{\vm_2}(u) = p_{\vb+\vec{\gamma};\vm_2}(u).
\end{split}
\end{equation}

\subsection{The Composite Alternative Generated by A Simple One}\label{sec:constructed_alt}
We are mostly concerned with testing the null hypothesis $\nullhyp$ against simple alternative hypotheses of the form $\{Q\}$ for some distribution $Q$ on $\cU$ 
(see Appendix~\ref{app:composite} for a discussion on how this can be extended to composite alternatives).
As stated earlier, throughout this paper we make the (weak) assumption that, for any considered alternative $Q$, there exists a $\vm^* \in \meanspace_{\nul}$ such that ${\mathbb E}_{\g{X} \sim Q}[\g{X}] = \vm^*$. 
By a standard property of exponential families, the KL divergence from $Q$ to $\nullhyp$ is then minimized by the element of $\nullhyp$ with the same mean as $Q$.
If~\eqref{eq:riprminb} holds, then $P_{\vm^*}$ must therefore be the reverse information projection of $Q$.
%
In that case, the GRO e-variable is the likelihood ratio between $Q$ and $P_{\vm^*}$, that is,
\begin{equation}
  \label{eq:simple}
  \frac{q(U)}{p_{\vm^*}(U)},
\end{equation}
i.e. it is a simple e-variable relative to $Q$. 
We will frequently use the fact 
(which follows from
Corollary 1 of \cite[Theorem 1]{GrunwaldHK23}) that there can be at most one simple e-variable with respect to any fixed alternative, i.e. of the form \eqref{eq:simple}. 
This is captured by the following proposition.
\begin{proposition}\label{prop:GRO}
    Fix a probability measure $Q$ on $U$. If there exists a simple e-variable relative to $Q$, then it must be the GRO e-variable for testing $\nullhyp$ against alternative $\{Q \}$. 
\end{proposition}
In particular, we will consider distributions $Q$ that admit a moment generating function and that have a density $q$ relative to the underlying measure $\nu$. 
While the former is a strong condition, it holds in many cases of interest. 
For our analysis, it will be beneficial to define a second exponential family $\althyp$ for $\g{U}$ with distributions $Q_{\vb; \vm^*}$ and corresponding densities 
\begin{equation}
    \label{eq:qfamily}
q_{\vb;\vm^*}(u) =  \frac{1}{Z_{\alt}(\vb; \vm^*)} \cdot \exp\left({\sum_{j=1}^d \beta_j t_j(u) }\right) \cdot  q(u),
\end{equation}
where $\vm^*$ is the mean of $X$ under $Q$, and $Z_{\alt}(\vb;\vm^*)$ is the normalizing constant.
The notational conventions that we use for $\althyp$ will be completely analogous to that for $\nullhyp$, e.g. $\vb_\alt(\cdot, \mu^*),$ $\vm_\alt(\cdot, \mu^*),$ $ \Sigma_\alt$, etc.
Since  $Q$ is assumed to have a moment generating function, the canonical domain $\canspace_{\alt, \vm^*}$ is nonempty and contains a neighborhood of $\vec{0}$.
Similarly, the mean-value space $\meanspace_{\alt}$ is also nonempty and contains a neighborhood of $\vm^*$. 
We further have the following: if we take any other  $Q' \in \althyp$, say $Q'= Q_{\vm'}$ for $\vm'\in \meanspace_\alt$, then the `constructed' family around $Q'$, i.e. $\{ q_{\vb; \vm'}:\vb \in \canspace_{\alt; \vm'}\}$ coincides with $\althyp$ (as was the case for the original $Q$, by~\eqref{eq:lin_rela}).

We may think of the null $\nullhyp$ and the generated family $\althyp$ as two different exponential families that share the same sufficient statistic. 
Moreover, as we shall see below, there are many examples where their mean-value spaces are equal, that is, $\meanspace_{\alt} = \meanspace_{\nul}$. 
In this case $\nullhyp$ and $\althyp$ are ``matching'' pairs:
they share the same sufficient statistic as well as the same set of means for this statistic. 
However, we stress that $\althyp$ is, first and foremost, a tool that is used in our analysis to help prove the existence of a simple e-variable with respect to $Q$.
It is only in special cases that $\althyp$ itself can be considered as (a subset of) the alternative (see e.g. Section~\ref{sec:regression} and Appendix~\ref{app:composite}).

\section{Existence of Simple Local E-Variables}
\label{sec:simpleH1general}
Here we will show how the family $\althyp$ is related to the question of whether $q(\g{U})/p_{\vm^*}(\g{U})$ is a {\em local\/} GRO e-variable around $\vm^*$. 
We say that a nonnegative statistic $S(U)$ is a local e-variable around ${\vm^*}$ if there exists a connected  open subset  $\canspace'_{\vm^*}$ of $\canspace_{\nul; \vm^*} \cap \canspace_{\alt; \vm^*}$ containing ${\bf 0}$ such that $S$  is an e-variable relative to $\nullhyp'= \{P_{\vb}: \vb \in \canspace'_{\vm^*} \}$, i.e.  $\sup_{\vb \in \canspace'_{ \vm^*}} \mathbb{E}_{P_{\vb; \vm^*}} [S] \leq 1$. 
If $S$ also maximizes $\mathbb{E}_Q[\ln S(U)]$ among all e-variables relative to $\nullhyp'$, then we say that $S$ is a local GRO e-variable with respect to $Q$.
A local (GRO) e-variable may not be an e-variable relative to the full null hypothesis $\nullhyp$, but it is a an e-variable relative to some smaller null hypothesis, restricted to all distributions in the null with mean in a neighborhood of $\vm^*$.
Investigating when local e-variables exist provides the basic insight on top of which the subsequent, much stronger Theorem~\ref{thm:main} about `global' e-variables is built. 
As stated in Section~\ref{sec:constructed_alt}, we may view $\nullhyp$ and $\althyp$ as two families with the same sufficient statistic, only differing in their carrier, which for $\nullhyp$ is  ${p}_{\vm^*} = p_{\vec{0}; \vm^*}$ and for $\althyp$ is $q_{\vec{0};  \vm^*} = q = q_{\vm^*}$: we can and will denote the original $Q$ also by $Q_{\vm^*}$. 

Define the function $f(\cdot;\vm^*):\canspace_{\nul; \vm^*} \cap \canspace_{\alt; \vm^*}\to \mathbb{R}$ as
\begin{equation}\label{eq:logZdifference}
f(\vb;\vm^*) := \log {\mathbb E}_{P_{\vb; \vm^*}}\left[\frac{q_{\vm^*}(\g{U})}{p_{\vm^*}(\g{U})}\right] 
= \log Z_{\alt}(\vb ; \vm^*) - \log Z_{\nul}(\vec\vb ; \vm^*),
\end{equation}
where the equality comes from the fact that we can rewrite the density in the numerator as $q_{\vm^*}(\g{U})=Z_q(\vb;\mu^*) \exp(\sum_{j=1}^d \beta_j t_j(u))^{-1} q_{\vb;\vm^*}(\g{U})$ and similar for the density in the denominator. 
It should be clear that the function $f(\cdot;\vm^*)$ is highly related to the question we are interested in.
Indeed, $q_{\vm^*}(\g{U})/p_{\vm^*}(\g{U})$ is a local e-variable relative to $\nullhyp' = \{P_{\vb}: \vb \in \canspace'_{\vm^*} \}$ if and only if $\sup_{\vb \in \canspace'_{\vm^*}} f(\vb;\vm^*) \leq 0$.
Equivalently, since $f(\mathbf{0};\vm^*)=\mathbf{0}$, we have that $q_{\vm^*}/p_{\vm^*}$ is a local e-variable around $\vm^*$ if and only if there  is a local maximum at $\mathbf{0}$.
To investigate when this happens, a standard result on exponential families gives the following:
\begin{equation} \label{eq:magic}
   \nabla f(\vb;\vm^*) = {\mathbb E}_{Q_{\vb;\vm^*}}[X] - 
   {\mathbb E}_{P_{\vb;\vm^*}}[X] 
\end{equation} 
In particular, it follows that $\nabla f({ \bf 0};\vm^*)  = \vm^*-\vm^*= {\bf 0}$.
Thus, $q_{\vm^*}/p_{\vm^*}$ is a local e-variable around $\vm^*$ if and only if the $d \times d$ Hessian matrix of second partial derivatives of $f(\cdot;\vm^*)$, is negative semidefinite in $\mathbf{0}$.
By \eqref{eq:logZdifference}-\eqref{eq:magic} and using a convex duality property of exponential families, this is equivalent to 
$$
I_{\nul}({\bf 0}; \vm^*) - I_{\alt}({\bf 0}; \vm^*) = 
\Sigma_{\nul}(\vm^*) - \Sigma_{\alt}(\vm^*)
\text{\ is positive semidefinite},
$$
where $I_{\nul}$ and $I_{\alt}$ denote the Fisher information matrix in terms of the canonical parameter spaces of $\nullhyp$ and $\althyp$, respectively.
We have thus proven our first result: 
\begin{proposition}\label{prop:local}
$q_{\vm^*}(\g{U})/p_{\vm^*}(\g{U})$ is a local e-variable around $\vm^*$ (and therefore, by Proposition~\ref{prop:GRO}, a GRO local e-variable) if and only if $\Sigma_{\nul}(\vm^*) - \Sigma_{\alt}(\vm^*)$ is positive semidefinite.     
\end{proposition}
The surprising result that follows below essentially adds to this that, if for every $\vm^* \in \meanspace_\alt$, $q_{\vm^*}/p_{\vm^*}$ is a local e-variable, then also for every $\vm^*$, we have that $q_{\vm^*}/p_{\vm^*}$ is a full, global e-variable!

\section{Existence of Simple Global E-Variables (Main Result)}
\label{sec:main}
The theorem below gives eight equivalent characterizations of when a global GRO e-variable exists.
Not all characterizations are equally intuitive and informative: the simplest ones are Part 1 and 3. 
To appreciate the more complicated characterizations as well, it is useful to first recall some convex duality properties concerning derivatives of KL divergences with regular exponential families~\cite[see e.g.][Section 18.4.3]{grunwald2007minimum}: 
\begin{align}\label{eq:betadiff}
 &   \vb_{\nul}(\vec{\mu} ; \vm^*) = \nabla_{\vm}  D(P_{\vec{\mu}} \| P_{\vm^*} ) ,
 \\  \label{eq:sigmadiff}
&  \left( \Sigma^{-1}_{\nul}(\vm) \right)_{ij} = \frac{d^2 }{d\vm_i d \vm_j} D(P_{\vec{\mu}} \| P_{\vm^*} ), 
\end{align} 
and analogous for $\althyp$.
That is, the gradient of the KL divergence in its first argument at $\vm$ is given by the canonical parameter vector corresponding to $\vm$, and the Hessian is given by the Fisher information, i.e. the inverse covariance matrix. 

\begin{theorem}\label{thm:main}
Let $\nullhyp$ be a regular exponential family with mean-value parameter space $\meanspace_{\nul}$. Fix a distribution $Q$ for $U$ with  ${\mathbb E}_{Q}[X] = {\vm}^*$ for some $\vec{\mu}^* \in \meanspace_{\nul} \subseteq \reals^d$ and consider the corresponding $\althyp$ as defined above via (\ref{eq:qfamily}). Suppose that $\meanspace_\alt$ is convex, $\meanspace_{\alt} \subseteq \meanspace_{\nul}$, and $\canspace_{\nul;\vm}\subseteq \canspace_{\alt;\vm}$ for all $\vm\in \meanspace_\alt$. Then the following statements are equivalent:
\begin{enumerate}
\item\label{item:sigmas} $\Sigma_{\nul}(\vm) - \Sigma_{\alt}(\vm)$ is positive semidefinite for all $\vm \in \meanspace_\alt$. 
\item\label{item:betas} $\left(\vb_{\nul}(\vm;\vm') - \vb_{\alt}(\vm;\vm')\right)^T \cdot  (\vm - \vm')\leq 0$ for all $\vm, \vm' \in \meanspace_\alt$.
\item\label{item:KL} $D(Q_{\vm}\| Q_{\vm'}) \geq D(P_{\vm}\| P_{\vm'})$ for all $\vm, \vm' \in \meanspace_\alt$. 
\item\label{item:norm_consts} $\log Z_{\nul}(\vb; \vm) \geq \log Z_{\alt}(\vb; \vm)$ for all $\vm \in \meanspace_\alt, \vb \in \canspace_{\nul; \vm}$. 
\item $q_{\vec{\mu}}(\g{U})/p_{\vec{\mu}}(\g{U})$ is a global e-variable for all  $\vec{\mu} \in \meanspace_{\alt}$.
\item $q_{\vec{\mu}}(\g{U})/p_{\vec{\mu}}(\g{U})$ is the global GRO e-variable w.r.t. $Q_\vm$ for all  $\vec{\mu} \in \meanspace_{\alt}$.
\item $q_{\vec{\mu}}(\g{U})/p_{\vec{\mu}}(\g{U})$ is a local e-variable for all  $\vec{\mu} \in \meanspace_{\alt}$.
\item $q_{\vec{\mu}}(\g{U})/p_{\vec{\mu}}(\g{U})$ is a local GRO e-variable w.r.t. $Q_\vm$ for all  $\vec{\mu} \in \meanspace_{\alt}$.
\end{enumerate}
\end{theorem}
Note that the canonical parameter space of a full exponential family is always convex, but the mean-value space need not be \citep{efron_2022}. Still, in all examples we consider below, the constructed family $\cQ$ will in fact be a {\em regular\/} exponential family, and then the convexity requirement must hold.

In the one-dimensional case, the first statement simplifies to  $\sigma_{\nul}^2(\mu) \geq \sigma_{\alt}^2(\mu)$ for all $\mu \in \meanspace_{\alt}$.
Similarly, the second statement reduces to  $\vb_{\alt}(\mu; \mu') \geq \vb_{\nul}(\mu; \mu')$ for all $\mu \in \meanspace_\alt$ such that $\mu > \mu'$ and  $\vb_{\alt}(\mu; \mu') \leq \vb_{\nul}(\mu; \mu')$ for all  $\mu \in \meanspace_\alt$ such that $\mu < \mu'$ for all $\mu' \in \meanspace_\alt$. 

Using standard properties of Loewner ordering, it can be established that $\Sigma_{\nul}(\vm) - \Sigma_{\alt}(\vm)$ is positive semidefinite if and only if $\Sigma^{-1}_{\alt}(\vm) - \Sigma^{-1}_{\nul}(\vm)$ is \citep[see e.g.][]{Agrawal18}.
Therefore, recalling (\ref{eq:betadiff}) and (\ref{eq:sigmadiff}), statement~\ref{item:sigmas} in Theorem~\ref{thm:main} can be thought of as a condition on the second derivative of $D(P_{\vm} \| P_{\vm^*}) - D(Q_{\vm} \| Q_{\vm^*})$, whereas statement~\ref{item:betas} refers to its first derivative, and statement~\ref{item:KL} to the difference in KL divergence itself. 
It is somewhat surprising that signs of differences between the second derivatives  and separately signs of differences between the first derivatives are sufficient to determine signs of difference between a function itself. 

\subsection{Simplifying Situations}
In some special situations, the conditions needed to apply Theorem~\ref{thm:main} may be significantly simplified.  We now identify two such situations, embodied by Proposition~\ref{prop:onedim} and Corollary~\ref{cor:fulldim}, that will be useful for our examples below. 

First, we note that it is sometimes easy to check that either $\meanspace_{\nul}=\meanspace_\alt$ or $\canspace_{\nul; \vec{\mu}^*}=\canspace_{\alt; \vec{\mu}^*}$. The following proposition shows that, in the 1-dimensional setting, this is already sufficient to apply the theorem (we do not know whether an analogous result holds in higher dimensions):
\begin{proposition}\label{prop:onedim}
Let $\nullhyp$ be a 1-dimensional regular exponential family with mean-value parameter space $\meanspace_{\nul}\subseteq \reals$. 
Fix a distribution $Q$ for $U$ with  ${\mathbb E}_{Q}[X] = \mu^*$ 
for some ${\mu}^* \in \meanspace_{\nul} $ and consider the corresponding $\althyp$ as defined above via (\ref{eq:qfamily}). 
Suppose that for all $\mu \in \meanspace_{\alt}$, we have $\sigma^2_p(\mu) \geq \sigma^2_q(\mu)$, i.e. the first condition of Theorem~\ref{thm:main} holds. Then:
\begin{enumerate}
    \item If $\meanspace_\alt = \meanspace_{\nul}$ then for all $\mu' \in \meanspace_{\alt}$,  $\canspace_{\nul;\mu'} \subseteq \canspace_{\alt ; \mu'}$, i.e., Theorem~\ref{thm:main} is applicable. 
    \item If for some $\mu \in \meanspace_{\alt}$, we have that $\canspace_{\nul; \mu} = \canspace_{\alt; \mu}$ then $\meanspace_\alt \subseteq  \meanspace_{\nul}$. Hence if for all $\mu \in \meanspace_{\alt}$, we have that $\canspace_{\nul; \mu} = \canspace_{\alt; \mu}$, then Theorem~\ref{thm:main} is applicable. 
\end{enumerate}
\end{proposition}
The proof is simple and we only sketch it here: for part 1, draw the graphs of $\beta_p(\mu;\mu')$ and $\beta_q(\mu;\mu')$ as functions of $\mu \in \meanspace_{\alt}$, noting that both functions must take the value $0$ at the point $\mu = \mu'$. Using that $1/\sigma^2_p(\mu)$ is the derivative of $\beta_p(\mu;\mu')$ and similarly for $\sigma^2_q(\mu)$, the function $\beta_q(\mu;\mu')$ must lie above  $\beta_p(\mu;\mu')$ for $\mu > \mu'$, and below for $\mu < \mu'$. Therefore the co-domain of $\beta_q$ must include that of $\beta_p$. The second part goes similarly,  essentially by flipping the just-mentioned graph of two functions by 90 degrees. 

Second, we note that, in practice, we often have a composite alternative $\cH_1$ in mind.
In such cases, Theorem~\ref{thm:main} can be used to determine whether there is a simple e-variable for each $Q\in \cH_1$.
If this is the case, an e-variable with respect to the entire alternative can be constructed using the method of mixtures, that is, by taking a convex mixture of all the simple e-variables (see Appendix~\ref{app:happyreviewer}).
In general, this process requires the construction of an auxiliary exponential family $\althyp$ and verifying one of the equivalent statements for each $Q\in \cH_1$, which can be quite tedious.
However, it turns out that it is often the case that $\cH_1$ coincides with the union of the set of families $\cQ$ that can be constructed from $\cP$  and any $Q \in \cH_1$.
That is, the auxiliary families that are constructed for different elements of $\cH_1$ might coincide; 
this is the case in the examples of Sections~\ref{sec:zero},~\ref{sec:moreksample},~\ref{sec:gausscale} and~\ref{sec:regression}. 
If this happens, it is sufficient to check the (pre)conditions of Theorem~\ref{thm:main} for each distinct family $\cQ$ that can be constructed rather than for each element of $\cH_1$.
This immediate corollary of Theorem~\ref{thm:main} is formally stated below. 
We will only explicitly invoke it in Section~\ref{sec:regression}, and while in that section, $\cH_1$ will itself be an exponential family, we stress that, in general, this need not be the case: to apply the corollary, it is sufficient for $\cH_1$ to be a {\em union\/} of exponential families. 
\begin{corollary}\label{cor:fulldim}
Let  $\nullhyp$ be a $d$-dimensional regular exponential family as before with   mean-value parameter space $\meanspace_{\nul}$, and let $\cH_1 = \bigcup_{\theta \in \Theta} \cQ^{(\theta)}$ where each $\cQ^{(\theta)}$ is a $d$-dimensional regular exponential family with the same sufficient statistic as $\nullhyp$ and with mean-value parameter space $\meanspace_q^{(\theta)}$ and canonical parameter spaces $\canspace^{(\theta)}_{q;\vm}$ for $\vm \in \meanspace_q^{(\theta)}$. Suppose that, for each $\theta \in \Theta$, for each $Q \in \cQ^{(\theta)}$, the corresponding set $\cQ$ as constructed above via (\ref{eq:qfamily}) in terms of $\cP$ and $Q$, happens to be equal to $\cQ^{(\theta)}$ and satisfies the pre-condition of Theorem~\ref{thm:main}, i.e. $\meanspace_q^{(\theta)}$ is convex, $\meanspace_q^{(\theta)} \subseteq \meanspace_{\nul}$, and $\canspace_{\nul;\vm}\subseteq \canspace^{(\theta)}_{q;\vm}$ for all $\vm\in \meanspace_q^{(\theta)}$.  
Then we have, with $Q^{(\theta)}_{\vm}$ (density $q^{(\theta)}_{\vm}$) denoting the element of $\cQ^{(\theta)}$ with mean $\vm$, for all $\theta \in \Theta$: 
\begin{quote}
For all  $\vm \in \meanspace_q^{(\theta)}$:
$\frac{q^{(\theta)}_{\vm}(\g{U})}{p_{\vm}(\g{U})}$ 
is the global GRO e-variable w.r.t. $Q^{(\theta)}_{\vm}$ $\Leftrightarrow$\\
For all $\vm \in \meanspace_q^{(\theta)}$: $\Sigma_{\nul}(\vm) - \Sigma^{(\theta)}_q(\vm)$ is positive semidefinite. 
\end{quote}
Here $\Sigma_q^{(\theta)}(\vm)$ denotes the $d \times d$ covariance matrix of the element of $\cQ^{(\theta)}$ with mean-value parameter vector $\vm$. 
\end{corollary}
\section{Examples}\label{sec:examples}
In this section we discuss a variety of settings to which Theorem~\ref{thm:main} can be applied. 
In some cases, this gives new insights into whether simple e-variables exist, and in others it simply gives a reinterpretation of existing results. 
The examples are broadly divided in terms of the curvature of the function $f(\cdot;\vm^*)$, as defined in \eqref{eq:logZdifference}.
Instances where  $f(\cdot;\vm^*)$ is constant will be referred to as having `zero curvature', those with a constant second derivative as having `constant curvature', and `nonconstant curvature' otherwise.

\subsection{Zero Curvature}\label{sec:zero}
\subsubsection{Multivariate Gaussian Location with shared covariance, constrained null}
\cite{SpectorCL23} considered the setting with $\nullhyp$ the multivariate Gaussian location family with some given nondegenerate covariance matrix $\Sigma$ and with mean restricted to $\vm \in \{(\mu_1, \ldots, \mu_d) \in \reals^d: \mu_1, \ldots, \mu_{d_0}= 0 \}$ for some $1 \leq d_0< d$; intuitively, $\mu_{d_0+1}, \ldots, \mu_{d}$ can be thought of as nuisance parameters. The simple alternative $Q$ is defined to be any fixed  Gaussian distribution with the same covariance matrix $\Sigma$ and some arbitrary mean $\vm^* \in \reals^d$. An easy calculation shows that the  family $\althyp$, generated from $Q$ and $\nullhyp$ as in~\eqref{eq:qfamily}, is the Gaussian location family with again covariance matrix $\Sigma$ but now mean restricted to $\vm \in \{(\mu_1, \ldots, \mu_d) \in \reals^d: (\mu_1, \ldots, \mu_{d_0})=  (\mu^*_1, \ldots, \mu^*_{d_0})\}$. Calculating further we find that  $\nullhyp$ and $\althyp$ are exponential families with the same $d'$-dimensional covariance matrix $\Sigma'_p=\Sigma'_q = \Sigma'$, where $d'= d- d_0$ and with entries $(\Sigma')_{i,j} = \Sigma_{d_0+i,d_0+j}$. Consequently, $\Sigma'_p-\Sigma'_q$ is the zero matrix and 
condition~\ref{item:sigmas} of Theorem~\ref{thm:main} is verified, and the simple likelihood ratio
$q_{\mu^*}(U)/p_{\mu^*}(U)$ is the global GRO e-variable with respect to $Q_{\mu^*}$.
This result was obtained earlier by  \cite{SpectorCL23} via a direct calculation.
\subsubsection{Gaussian and Poisson k-sample tests}
\cite{HaoGLLA23} provide GRO e-values for $k$-sample tests with regular exponential families. 
In their setting, data arrives in $k\in\mathbb{N}$ groups, or samples, and they test the hypothesis that all of the data points are distributed according to the same element of some exponential family.  They show that in case this exponential family is the Poisson or Gaussian location family, then the resulting e-variables are simple. We now show how their results can be re-derived using our Theorem~\ref{thm:main}. 

Specifically, let $U=(Y_1,\dots,Y_k)$ for $Y_i\in \mathcal{Y}$, so that $\mathcal{U}=\mathcal{Y}^k$ for some measurable space $\mathcal{Y}$.
Furthermore, fix a one-dimensional regular exponential family on $\mathcal{Y}$, given in its mean-value parameterization as $\nullhyp_{\textsc{start}} = \{P_{\mu}: \mu \in \meanspace_{\textsc{start}}\}$ with sufficient statistic $t_{\textsc{start}}(Y)$.
The composite null hypothesis $\nullhyp$ considered in the k-sample test expresses that $Y_1,\dots,Y_k\stackrel{\mathrm{i.i.d.}}{\sim} P_{\mu}$ for some $\mu \in \meanspace_{\textsc{start}}$.
On the other hand, the simple alternative $Q$ that \citeauthor{HaoGLLA23} consider is characterized by $\vm=(\mu_1, \ldots, \mu_k) \in \meanspace_{\textsc{start}}^k$, and expresses that the $Y_1, \ldots, Y_k$ are independent with $Y_i \sim P_{\mu_i}$ for $i=1\dots k$. 
They show that, for the case that $\nullhyp_{\textsc{start}}$ is either the Gaussian location family or the Poisson family, 
$$
S(U):= \prod_{i=1}^k\frac{ p_{\mu_i}(Y_i)}{p_{\bar{\mu}}(Y_i)},\ \text{with\ }
\bar{\mu}=\frac1k \sum_{i=1}^k \mu_i,$$
is a simple e-value relative to $Q$, and that its expectation is constant as the null varies.
That is, for any $\mu'\in \meanspace_{\textsc{start}}$, it holds that 
\begin{equation}\label{eq:k-sample_estat}
    \mathbb{E}_{\g{U}\sim P_{\mu'}\times\dots\times P_{\mu'}}\left[S(U) \right] = 1.
\end{equation} 
It follows from Proposition~\ref{prop:GRO} that $S(U)$ must be the GRO e-variable for testing $\cP$ against $Q$.
This finding can now be re-interpreted as an instance of Theorem~\ref{thm:main}, as we will show in detail for the Poisson family; the analysis for the Gaussian location family is completely analogous.

%
In the Poisson case, $t_{\textsc{start}}(Y)=Y$, so that $\nullhyp$ defines an exponential family on $\mathcal{U}$ with sufficient statistic $X=\sum_{i=1}^k Y_i$ and mean-value space $\meanspace_\nul = \mathbb{R}^+$.
The latter follows because the sum of Poisson data is itself Poisson distributed with mean equal to the sum of means of the original data.
Under the alternative, the mean of the sufficient statistic is given by  $\mu^*:=\mathbb{E}_Q[\sum_{i=1}^k Y_i]=\sum_{i=1}^k \mu_i$, so that the elements of the auxiliary exponential family $\mathcal{Q}$ as in~\eqref{eq:qfamily} can be written as
\begin{equation}\label{eq:k-sample_altfam}
    \alt_{\beta;\mu^*}(Y_1,\dots,Y_k)= \frac{1}{Z_{\alt}(\beta;\mu^*)} \cdot \exp\left(\beta \sum_{i=1}^k Y_i\right) \cdot q(Y_1,\dots,Y_k).
\end{equation}
Note in particular that $\mathcal{Q}$ is, by construction, a one-dimensional exponential family with sufficient statistic $\sum_{i=1}^k Y_i$, which does not equal (yet may be viewed as a subset of) the full $k$-dimensional exponential family from which $Q$ was originally chosen.
The normalizing constant $Z_\alt(\beta;\mu^*)$ is equal to the moment generating function of $X$ under $Q$, which is given by
\[Z_\alt(\beta;\mu^*)=\mathbb{E}_Q\left[ \exp\left(\beta \sum_{i=1}^k Y_i\right)\right] = \exp\left(\mu^*(e^\beta -1)\right).\]
It follows that
\[\mathbb{E}_{Q_{\beta; \mu^*}}\left[ \sum_{i=1}^k Y_i\right] = \frac{\mathrm{d}}{\mathrm{d}\beta} \log Z_\alt(\beta;\mu^*) = \mu^*e^\beta ,\]
which shows that mean-value space of the alternative is again given by $\meanspace_\alt=\mathbb{R}^+$. 
Therefore, via Proposition~\ref{prop:onedim}, the assumptions of Theorem~\ref{thm:main} are satisfied. 
The element of $\nullhyp$ with mean $\mu^*$ is given by $P_{\bar{\mu}}\times \dots\times P_{\bar{\mu}}$, so that
\[\frac{\alt_{ \mu^*} (U)}{\nul_{\mu^*}(U)}=\prod_{i=1}^k\frac{ \nul_{\mu_i}(Y_i)}{\nul_{\bar{\mu}}(Y_i)}.\]
Under $P_{\mu^*}$, the sufficient statistic $\sum_{i=1}^k Y_i$ has the same distribution as under $Q_{\mu^*}$, so that $ Z_p (\beta;\mu^*)=Z_q(\beta; \mu^*)$.  
Consequently, $f(\cdot;\vm^*)$ as in $\eqref{eq:logZdifference}$ is zero, so that its second derivative is zero, and condition~\ref{item:sigmas} of Theorem~\ref{thm:main} is verified.
It follows that, $q_{\mu^*}(U)/p_{\mu^*}(U)$ is the global GRO e-variable with respect to $Q_{\mu^*}$.

\subsection{Constant Curvature: Multivariate Gaussian Location, distinct covariance}
\cite{hao2024expfamgeneral} considered, in great detail, the setting with $\nullhyp$ the full multivariate Gaussian location family with some given nondegenerate covariance matrix $\Sigma_{\nul}$ and let
$Q$ be any  Gaussian distribution with nondegenerate covariance matrix $\Sigma_{\alt}$. 
Note that in this case we have that $\g{X}=\g{U}$, i.e. the sufficient statistic is simply given by the original data.
The family $\althyp$, generated from $Q$ and $\nullhyp$ as in~\eqref{eq:qfamily}, is the full Gaussian location family with fixed covariance matrix $\Sigma_{\alt}$.
For both $\nullhyp$ and $\althyp$, the mean-value and canonical spaces are all equal to $\reals^d$, 
so that the regularity assumptions behind Theorem~\ref{thm:main} are satisfied.
Furthermore, the covariance functions are constant, since $\Sigma_{\nul}(\vm) = \Sigma_{\nul}$ and $ \Sigma_{\alt}(\vm) = \Sigma_{\alt}$ for all $\vm \in \reals^d$.
It follows that, if $\Sigma_{\nul} - \Sigma_{\alt}$ is positive semidefinite, then $\Sigma_{\nul}(\vm)-\Sigma_{\alt}(\vm)$ is positive semidefinite for all $\vm \in \reals^d$.
In that case, Theorem~\ref{thm:main} shows that the simple likelihood ratio $q_{\vm}/p_{\vm}$ is the GRO e-value w.r.t. $Q_{\vm}$ for every $\vm \in \reals^d$. 
The growth rate is given by
$$
\mathbb{E}_Q \left[ \log \frac{q_{\vm}(U)}{p_{\vm}(U)}\right] = D_{\gauss}(\Sigma_{\alt} \Sigma_{\nul}^{-1}),
$$
where $D_{\gauss}(B) := \frac{1}{2} \left( - \log \det(B) - \left( d - \trace(B) \right)   \right)$, i.e. the standard formula for the KL divergence between two multivariate Gaussians with the same mean. 

In the case that  $\Sigma_{\nul} - \Sigma_{\alt}$ is negative semidefinite, the simple likelihood ratio does not give an e-value.
However, the GRO e-value for this case can still be derived, as shown by~\citet[Part 4 of Theorem~1]{hao2024expfamgeneral}.

\subsection{Nonconstant Curvature: Univariate Examples}
We now discuss three examples with nonconstant curvature. In the first two,  Theorem~\ref{thm:main} can be used to show the existence of simple e-variables.
All three are univariate in nature; in the separate Section~\ref{sec:regression} we provide the example of linear regression, which has nonconstant curvature but is multivariate. 
 
\subsubsection{More k-Sample Tests}\label{sec:moreksample}
Consider again the $k$-sample test setting of Section~\ref{sec:zero}. 
Besides the Gaussian and Poisson case, \cite{HaoGLLA23} identify one more model that gives rise to a $k$-sample test with a simple e-value: the case that $\nullhyp_{\textsc{start}}$ is the Bernoulli model.
The difference with the Gaussian location- and Poisson family is that the involved e-value does not have constant expectation 1 here. 
Nevertheless, this result for the Bernoulli model can also be cast in terms of Theorem~\ref{thm:main} using a different argument. 

\begin{figure}[ht]
    \centering
    \includegraphics[width=0.5\textwidth]{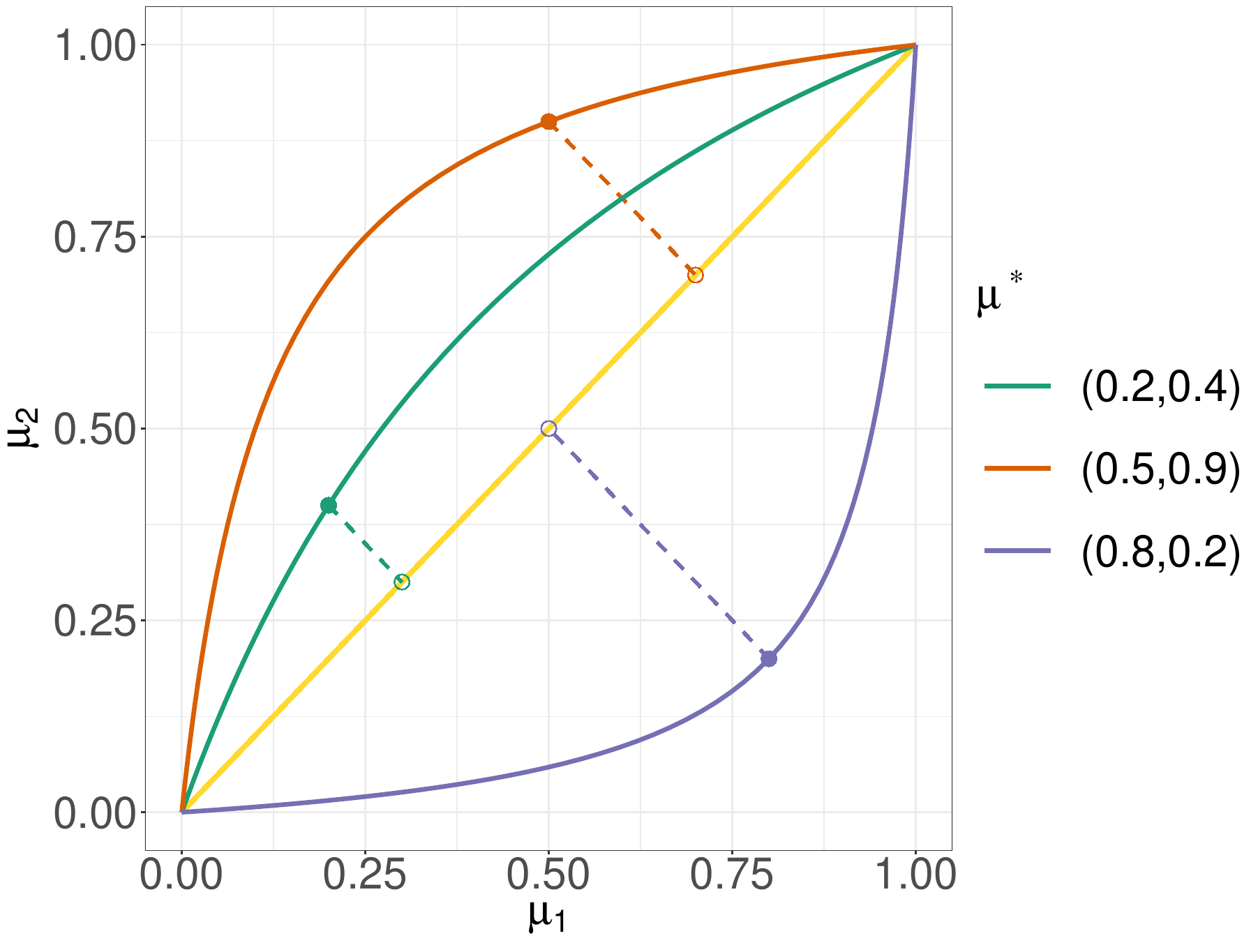}
    \caption{The family $\althyp$ for various $\vm^*$. The coordinate grid represents the parameters of the full $2$-sample Bernoulli family, the straight line shows the parameter space of $\nullhyp$, the curved lines show the parameters of the distributions in $\althyp$, and the dashed lines show the projection of $\vm^*$ onto the parameter space of $\nullhyp$.}
    \label{fig:bernoulliQfam}
\end{figure}

Again, $\nullhyp$ is an exponential family on $\mathcal{U}$ that states that the $k$ samples are i.i.d. Bernoulli, which has sufficient statistic $X=\sum_{i=1}^k Y_i$.
Its mean-value space is given by $\meanspace_\nul = (0, k)$, since the sum of $k$ i.i.d. bernoulli random variables with parameter $\mu$ has a binomial distribution with parameters $(k,\mu)$. 
Under the alternative $Q$, the $k$ samples are independently Bernoulli distributed with means given by $\vm \in (0,1)^k$, in which case the sum has mean $\mu^*=\sum_{i=1}^k \mu_i$.
When constructing the family $\althyp$ as in~\eqref{eq:qfamily}, it can be verified that $Q_{\beta,\mu^*}$ is the product of Bernoulli distributions with means 
\begin{equation}\label{eq:alt_ber_means}
    \left(\frac{e^\beta \mu_1}{1-\mu_1+e^\beta \mu_1},\dots,\frac{e^\beta \mu_k}{1-\mu_k+e^\beta \mu_k}\right).
\end{equation}
This family of distributions is illustrated in Figure~\ref{fig:bernoulliQfam} for different choices of $\vm^*$.
Seen as a function of $\beta$, all entries in~\eqref{eq:alt_ber_means} behave as sigmoid functions, so that the sum takes values in $(0,k)$. 
It follows that the mean-value space of $\althyp$ is given by $\meanspace_\alt = (0,k)$, which equals $\meanspace_\nul$ --- and also, the canonical spaces are all equal to $\reals$.
Furthermore, the normalizing constant $Z_q(\beta;\mu^*)$ of $\althyp$ must be given by
\[Z_\alt(\beta;\mu^*)=\prod_{i=1}^k (1-\mu_i+\mu_i e^\beta).\]

We will now verify that item~\ref{item:norm_consts} of Theorem~\ref{thm:main} is satisfied by doing a similar construction for arbitrary $\mu \in (0,k)$.
The element in $\nullhyp$ with mean $\mu$ corresponds to Bernoulli parameter $\mu/k$, so that we have
\[Z_\nul(\beta;\mu)=\mathbb{E}_{P_{\mu^*}}\left[\exp\left(\beta \sum_{i=1}^k Y_i\right) \right]= \left(1-\frac{\mu}{k}+\frac{\mu}{k}e^\beta \right)^k.\]
Furthermore, there is a corresponding $\vm'\in (0,1)^k$ such that  $\sum_{i=1}^k \mu'_i=\mu$ and $\vm'$ can be written as~\eqref{eq:alt_ber_means} for a specific $\beta$. 
Repeating the reasoning above gives
\[Z_\alt(\beta;\mu)=\prod_{i=1}^k (1-\mu'_i+\mu'_i e^\beta).\]
By concavity of the logarithm, it holds that
\[\log Z_\nul(\beta;\mu)=k\log \left(1-\frac{\mu}{k}+\frac{\mu}{k}e^\beta \right) \geq \sum_{i=1}^k \log(1-\mu_i'+\mu_i'e^\beta)=\log Z_\alt(\beta;\mu).\]
We can therefore conclude that $q(U)/p_{\mu^*}(U)$ is the GRO e-variable with respect to $Q$.
The family $\cQ$ constructed coincides with the alternative $\cH_1$ one would consider when testing for a minimal {\em odds ratio\/} effect size, which is indeed often considered in practice. In Appendix~\ref{app:happyreviewer} we contrast this with an alternative choice of $\cH_1$ based on another popular notion of effect size.

\cite{HaoGLLA23} investigate several other exponential families for $k$-sample testing, such as exponential distributions, Gaussian scale, and beta, but none of these give rise to a simple e-value.
Parts 1-4 of Theorem~\ref{thm:main} provide some insight into what separates these families from the Gaussian location, Poisson, and Bernoulli.

\subsubsection{Gaussian Scale Family}\label{sec:gausscale}
Another setting in which Theorem~\ref{thm:main} applies is where $\nullhyp$ equals the Gaussian scale family with fixed mean, which we take to be $0$ without loss of generality. 
That is, $\nullhyp = \{ P_{\sigma^2}: \sigma^2 \in \meanspace_{\nul} \}$ where $P_{\sigma^2}$ is the normal with mean $0$ and variance $\sigma^2$, i.e.
\begin{equation}
 \label{eq:gaussvar}
p_{\sigma^2}(U) = \frac{1}{\sqrt{2 \pi} \sigma} \cdot 
e^{-\frac{1}{2\sigma^2} U^2}.
\end{equation}
We will substantially extend this null hypothesis, and hence this example, in Section~\ref{sec:regression}.
For now, note that $\nullhyp$ is an exponential family with sufficient statistic $X=U^2$, mean-value parameter $\sigma^2$ and mean-value space given by $\meanspace_\nul = \mathbb{R}^+$. 
The canonical parameterization of the null relative to any mean-value $\sigma^{2}\in \meanspace_\nul$ is given by
\begin{equation}\label{eq:gaussvarb}
p_{\beta; \sigma^{2}}(U) 
= \frac{1}{Z_{\nul}(\beta; \sigma^{2})} \cdot 
e^{\beta U^2} \cdot \frac{1}{\sqrt{2 \pi \sigma^{2}}} e^{- U^2/ (2\sigma^{2} )}
\end{equation}
with canonical parameter space $\canspace_{\nul; \sigma^{2}} = (- \infty, 1/(2 \sigma^{2}))$.

As alternative, we take $Q$ to be a Gaussian distribution with some fixed mean $m \neq 0$ and variance $s^2$. 
We use $m$ and $s^2$ instead of $\mu$ and $\sigma^2$ here to avoid confusion with the mean-value parameters of $\nullhyp$.
The expected value of $X$ under $Q$ is given by $\sigma^{*2}:=\mathbb{E}_Q[X]=s^2+ m^2$.
The family $\althyp = \{Q_{\beta}: \beta \in \canspace_{\alt; \sigma^{*2}} \}$ as defined by \eqref{eq:qfamily} therefore becomes: 
\begin{equation}\label{eq:altfam_gaussscale}
q_{\beta; \sigma^{*2} }(U)  = \frac{1}{Z_{\alt}(\beta; \sigma^{*2})} \cdot 
e^{\beta U^2 } \cdot \frac{1}{\sqrt{2 \pi} s} \cdot 
e^{- c(U- m)^2}, \     
\end{equation}
where $c = 1/(2 s^2)$, with $\canspace_{\alt; \sigma^{*2}}  = (- \infty,  c )$.
Comparing (\ref{eq:gaussvarb}) and the above confirms that $\althyp$ is an exponential family that has the same sufficient statistic, namely $U^2$, as $\nullhyp$, but different carrier.  

The normalizing constant $Z_{\alt}$ can be computed using (for example) the moment generating function of the noncentral chi-squared.
 \begin{align*}
     Z_\alt(\beta;\sigma^{*2}) &= \mathbb{E}_Q\left[ e^{\beta U^2} \right]
     = \mathbb{E}_Q\left[ e^{\beta s^2 (\frac{U}{s})^2}\right]
     = (1-2\beta s^2)^{-1/2} \exp\left(\frac{m^2 \beta}{1-2\beta s^2}\right),
 \end{align*}
where we use that $(U/s)^2$ has noncentral chi-squared distribution with one degree of freedom and noncentrality parameter $m^2/s^2$.
Plugging this back in~\eqref{eq:altfam_gaussscale} shows that $q_{\beta,\sigma^{*2}}$ is a normal density with mean $c m /(c-\beta)$ and variance $1/(2 (c-\beta)) = s^2/(1- 2 \beta s^2)$. 
This gives
\begin{align}\label{eq:gauss_second_moment}
{\mathbb E}_{Q_{\beta; \sigma^{*2}}}[U^2] 
=  \frac{2c^2 m^2 -(\beta-c)}{2(\beta-c)^2}
\end{align}
The mean-value parameter space of $\althyp$ is thus given by $\meanspace_{\alt} = 
\{ {\mathbb E}_{Q_{\beta; \sigma^{*2}}}[U^2], \beta < c \}  = \reals^+
$ which is equal to  $\meanspace_{\nul}$. 
Thus, this constructed family does not equal the natural choice of composite alternative that $Q$ was also chosen from, i.e. the (two-dimensional) set of all Gaussians with arbitrary variance mean unequal to zero. 
However, it does correspond to a specific one-dimensional subset thereof, as was illustrated in Figure~\ref{fig:gausscaleQfam} in the introduction.



Since $\meanspace_{\alt}= \meanspace_{p}$, we get, via Proposition~\ref{prop:onedim} that a simple e-variable w.r.t. $Q$ exists if, for all $\sigma^2 >0$, we have that 
$\textsc{var}_{P_{\sigma^2}}[U^2] \geq \textsc{var}_{Q_{\sigma^2}}[U^2]$. We now show this to be the case. We have 
\[\textsc{var}_{P_{\sigma^2}}[U^2] = 2 \sigma^4 = 2 (\mathbf{E}_{P_{\sigma^2}}[U^2])^2 = 2 (\mathbf{E}_{Q_{\sigma^2}}[U^2])^2.\]
It is therefore sufficient to check whether, for all $\sigma^2 > 0$, it holds that $
\textsc{var}_{Q_{\sigma^2}}[U^2] \leq 2 ({\mathbb E}_{Q_{\sigma^2}}[U^2])^2$.
We can either verify this  using existing results by noting that, no matter how $m$ and $s^2$ were chosen, $U^2$ has a noncentral $\chi^2$-distribution under each $Q_{\sigma^2}$, for which it is known that the inequality holds. 
We can also easily verify it explicitly  now that we have already found an expression for $Z_q(\beta; \sigma^{*2})$: since there is no more mention of the null hypothesis, it is equivalent to check whether for each $\beta \in \canspace_{\alt; \sigma^{*2}}$ we have 
\[\textsc{var}_{Q_{\beta; \sigma^{*2}}}[U^2] \leq  2 \left( \mathbb{E}_{Q_{\beta,\sigma^{*2}}}[U^2] \right)^2.\]
To this end, the variance function in terms of $\beta$ can be computed as 
\begin{equation}\label{eq:gauss_scale_var}
    \textsc{var}_{Q_{\beta; \sigma^{*2}}}[U^2]  = \frac{d^2}{d \beta^2} \log Z_{\alt}(\beta; \sigma^{*2}) = -\frac{4c^2 m^2 -(\beta-c)}{2(\beta-c)^3}.
\end{equation}
Comparing this to~\eqref{eq:gauss_second_moment} shows that the condition above indeed holds, from which we can conclude that $q(\g{U})/p_{\sigma^{*2}}(\g{U})$ is an e-value.

Finally, note that even though the mean-value parameter spaces of $\nullhyp$ and $\althyp$ are equal, the canonical spaces are not: $\canspace_{\nul;  \sigma^{*2}}$ is a proper subset of $\canspace_{\alt; \sigma^{*2}}$.
More generally, for any $\sigma'^2>0$ different from the $\sigma^{*2}$ we started with, the canonical spaces  $\canspace_{\nul;  \sigma'^{2}}$  and $\canspace_{\alt; \sigma'^{2}}$ both change but remain unequal. 
Still, Proposition~\ref{prop:onedim} ensures that we will have  $\canspace_{\nul;  \sigma'^{2}} \subset \canspace_{\alt; \sigma'^{2}}$.

\subsubsection{NEFS and their Variance Functions}\label{sec:nef}
In this section, we consider the setting where $\nullhyp$ is a one-dimensional natural exponential family (NEF) and $Q$ is also an element of an NEF.
This setting is particularly suited for the analysis above, because the constructed family $\althyp$ can be seen to equal the NEF that $Q$ was chosen from.
We therefore do not differentiate between the simple or composite alternative in this section.
Furthermore, NEFs are fully characterized by the pair $(\sigma^2(\mu),\meanspace)$, where $\meanspace$ is the mean-value parameter space and $\sigma^2(\mu)$ is the variance function as defined before. 
A wide variety of NEFS and their corresponding variance functions have been studied in the literature~\cite[see e.g.][]{morris1982natural,Jorgensen97,BarlevLR24} and this can be used in conjunction with Theorem~\ref{thm:main} to quickly check on a case-by-case basis whether any given pair of NEFs provides a simple e-variable. 

For example, let  $\nullhyp = \{P_{\lambda,r}: \lambda \in  \reals^+ \}$ be the set of Gamma distributions for $U$ with varying scale parameter $\lambda$ and fixed shape parameter $r>0$.
The sufficient statistic is given by $X=U$ and its mean under $P_{\lambda,r}$ equals $r \lambda$, so the mean-value parameter space is $\meanspace_\nul=\reals^+$. 
The variance function is given by $\sigma^2_\nul(\mu)=\mu^2/r$.
If we set $Q$ to $P_{\lambda^*, r'}$ for specific $\lambda^*,r'\in \reals^+$, then $\althyp$ is the set of Gamma distributions with fixed shape parameter $r'$.

Similarly, let $\nullhyp$ be the set of negative binomial distributions with fixed number of successes $n\in \mathbb{N}$ and let $Q$ be any Poisson distribution, so that $\althyp$ equals the Poisson family.
The variance functions are given by $\sigma^2_\nul(\mu)=\mu^2/n+ \mu$ and $\sigma^2_\alt(\mu)=\mu$, respectively. 
It is trivially true that $\sigma^2_\nul(\mu)\geq \sigma^2_\alt(\mu)$ for all $\mu$, so Theorem~\ref{thm:main} reveals that a simple e-variables exists with respect to any element of the Poisson family. 
More generally, we may look at the Awad-Bar-Lev-Makov (ABM) class of NEFs~\citep{BarLevR21,awad2022new,bar2023exponential} that are characterized by mean-value parameter space $\meanspace=\reals^+$ and variance function
$$
\sigma^2_s(\mu) = \mu \left(1 +\frac{\mu}{s} \right)^r,\ s > 0,\ r=0, 1,2,... 
$$
This class was proposed as part of a general framework for zero-inflated, over-dispersed alternatives to the Poisson model (which would arise for $r=0$). 
The case $r=1$ recovers the negative binomial distribution and $r=2$ is called the generalized Poisson or Abel distribution.
As was the case for the negative binomial distribution, it follows from Theorem~\ref{thm:main} that simple e-variables exist for testing any of the ABM NEFs against the Poisson model.

Much more generally, consider the Tweedie-Bar-Lev-Enis class \citep{bar2020independent} of NEFs that have mean-value space $\meanspace = \reals^+$ and power variance functions
\[
\sigma^2(\mu)=a \mu^{\gamma },\ a>0,\ \mu>0,\ \gamma \geq 1. 
\]
We require $\gamma\geq 1$ because there are no families of this form with $\gamma \in (0,1)$ and while there are families in this class with $\gamma < 0$, they are not regular and therefore beyond the scope of this paper.
The cases $\gamma=1$ (Poisson) and $\gamma=2$ (Gamma families, with $a$ depending on the shape parameter) were already encountered above.  
If we test between two of such families, say $\nullhyp$ with $\sigma_{\nul}^2(\mu) = a_{\nul} \mu^{\gamma_{\nul}}$ and $\althyp$ with $\sigma_{\alt}^2(\mu) = a_{\alt} \mu^{\gamma_{\alt}}$  that share the same underlying sample space, there do not exist simple e-variables in general.
Indeed, we have that 
$\sigma^2_{\nul}(\mu) \geq \sigma^2_{\alt}(\mu)$ 
if and only if $\mu^{\gamma_\nul-\gamma_\alt} \geq a_\alt / a_\nul$, which, for certain combinations of parameters, does not hold for all $\mu \in \meanspace$.
Since this condition might hold for some $\mu$ but not for others, this suggests that there may be cases where we find local e-variables that are not global. 

\begin{figure}[ht]
    \centering
    \includegraphics[width=0.5\textwidth]{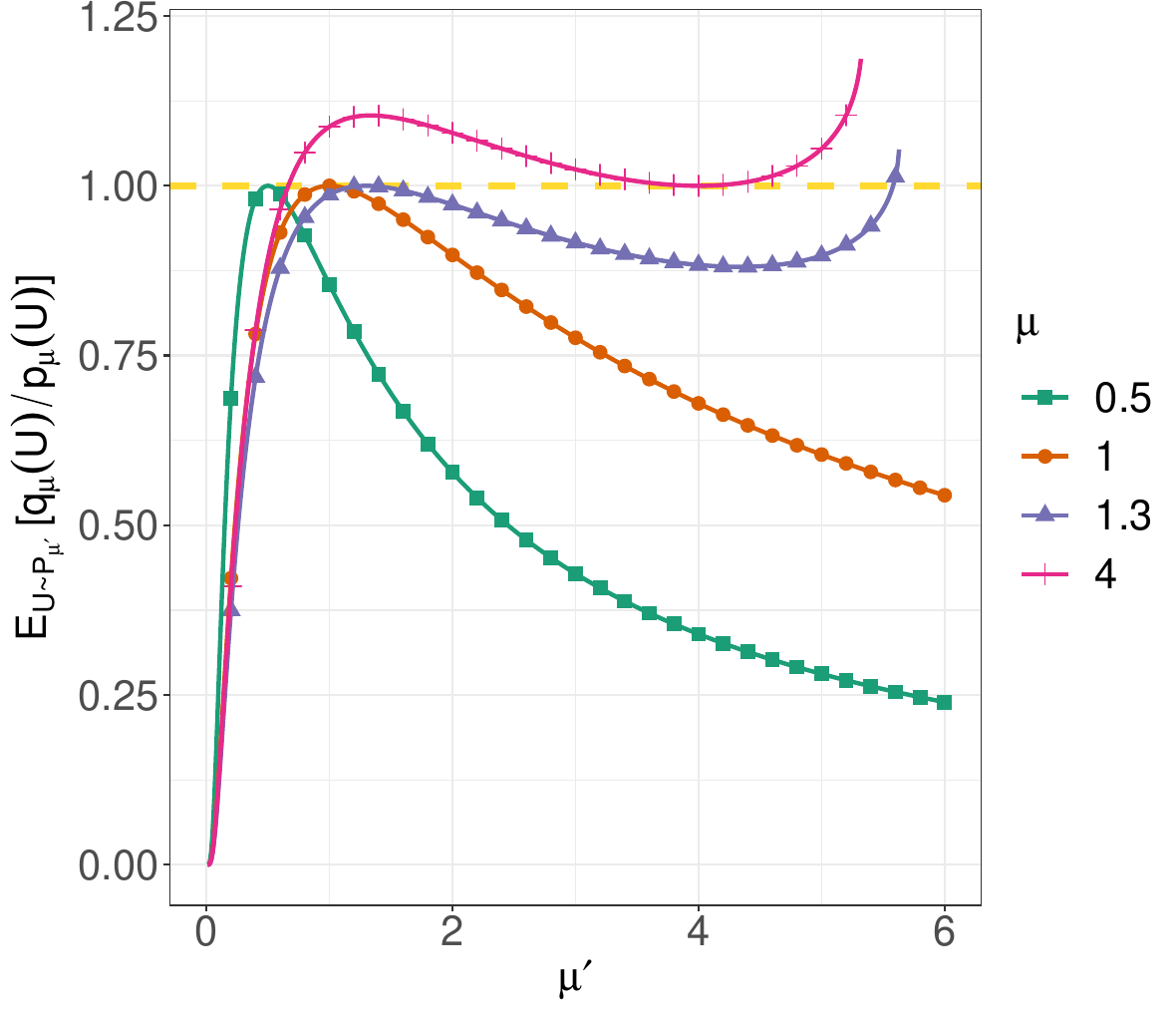}
    \caption{The expected value of $q_\mu(U)/p_\mu(U)$ under the null $P_{\mu'}$ for varying $\mu'$.}
    \label{fig:local_evar}
\end{figure}

Let us investigate this for $(a_\nul,\gamma_\nul)=(1,2)$ and $(a_\alt,\gamma_\alt)=(1/2,3)$, which corresponds to the family of exponential distributions and the family of inverse Gaussian distributions with shape parameter $\lambda:=a_\alt^{-1}=2$ respectively. 
In this case, it holds that $\sigma^2_{\nul}(\mu) \geq \sigma^2_{\alt}(\mu) \Leftrightarrow \mu \leq a_\alt^{-1}$. 
It follows from the analysis in Section~\ref{sec:simpleH1general} that $q_\mu(U)/p_\mu(U)$ is a local e-variable for $\mu \leq a_\alt^{-1}$.
However, since the condition does not hold for all $\mu$ we cannot use Proposition~\ref{prop:onedim} (or, equivalently, because, as we will see, the preconditions for Theorem~\ref{thm:main} do not hold), this need not necessarily also be a global e-variable.
In fact, the expected value under $\mu'\in \meanspace$ is given by
\begin{equation}\label{eq:exp_integral}
    \mathbb{E}_{U\sim P_{\mu'}}\left[\frac{q_\mu(U)}{p_\mu(U)}\right] = \int_0^\infty \frac{1}{\mu'} \sqrt{\frac{\lambda}{2\pi x^3}} \exp\left( -\frac{\lambda(x-\mu)^2}{2\mu^2 x} +\frac{x}{\mu}  -\frac{x}{\mu'}\right) dx ,
\end{equation} 
which diverges for $\mu'\geq (1/\mu -\lambda/(2\mu^2))^{-1}$. 
The latter is vacuous for $\mu \leq \lambda/2$, which means that for such $\mu$ we might still get a global  e-variable.
For $\mu \in (\lambda/2, \lambda)$, this shows that we will get a local e-variable that is not a global e-variable.
These different regimes are illustrated in Figure~\ref{fig:local_evar}.
For $\mu>1$, the lines stop when the integral in~\eqref{eq:exp_integral} starts diverging. 
To see how the potential divergence (for large enough $\mu'$, in the regime $1 < \mu < 2$) plays out in terms of the function $f$ in (\ref{eq:logZdifference}), consider for example $\mu=3/2$. Then, as is immediate from the definition of exponential distributions and the inverse Gaussian density with $\lambda=2$ we have $q_{\beta; \mu}(x) \propto \exp((\beta -4/9) x) h(x)$ with $h$ the probability density on $\reals^+$ given by $h(x)= \sqrt{1/(\pi x^3)} \exp(-1/x)$, whereas $p_{\beta;\mu} \propto \exp((\beta- 2/3) x)$. We see that $\canspace_{p;\mu} = (- \infty,6/9)$ whereas $\canspace_{q;\mu}= (-\infty, 4/9)$. Thus, as $\beta \uparrow 4/9$, we get that $\log Z_p(\beta)$  converges to a finite constant whereas $\log Z_q(\beta) \uparrow \infty$, so that $f(\beta,\mu) \rightarrow \infty$, with $f$ the function in (\ref{eq:logZdifference}), as it should. 

\subsection{The Linear Model}\label{sec:regression}
We now show that Theorem~\ref{thm:main} allows us to conclude that simple e-variables exist for the linear model, i.e. standard linear regression with Gaussian noise, where the null hypothesis $\cP$ is a subset of the alternative $\cH_1$ obtained by setting the regression parameter of a control random variable to $0$, if we allow the variance in $\cP$ to be a free parameter. 
This was shown directly, without associating a specific family $\althyp$ to $\cP$,  in an unpublished master thesis \citep{DeJong21}. De Jong's treatment involved a lot of hard-to-interpret calculus, much of it discovered by trial-and-error. The advantage of the present treatment is that Theorem~\ref{thm:main} clearly guides the reasoning and suggests what formulas to verify. The setting is really a vast extension of that of Section~\ref{sec:gausscale}, which is (essentially) retrieved if below we set $d=0$. 
Interestingly, e-variables for linear models were already derived by \cite{perez2024estatistics} and \cite{lindon2024anytime}, based on right-Haar priors. The current approach provides a different type of e-variable which has the advantage that it does not require the variance under the alternative to be equipped with a right-Haar prior: while for convenience we give the treatment below for $\cH_1$ with the variance $\sigma^2$ being left a free parameter, we can freely apply the results to any $\cH'_1 \subset\cH_1$, in particular with $\cH'_1$ restricted to densities with a fixed variance. The price to pay is that the e-variables derived below, while growth-optimal for the fixed $Q \in \cH_1$ relative to which they are defined, will in general not be GROW ({\em worst-case\/} growth optimal, see \citep{GrunwaldHK23}) in the worst-case over all distributions in $\cH_1$ when $\sigma^2$ varies within $\cH_1$. 

Assume then that data arrive as a block of outcomes together with given covariate vectors, i.e. $U=((Y_1,\vec{x}_1),\dots, (Y_n, \vec{x}_n))$ with $Y_i\in \reals$ and $\vec{x}_i = (x_{i,0}, x_{i,1}, \ldots, x_{i,d})^T \in \mathbb{R}^{d+1}$. 
In linear regression with Gaussian noise, it is assumed that each $Y_i$, conditionally on $\vec{x}_i$, is normally distributed with mean $\vec{\gamma}^T \vec{x}_i$ and variance $\sigma^2$, where $\vec{\gamma}\in \mathbb{R}^{d+1}$ and $\sigma^2\in (0,\infty)$ are parameters shared between observations.
We furthermore make the (standard) assumption that $n \geq d+1$ and that the matrix $(\vec{x}_1, \ldots, \vec{x}_n)$ has maximal (i.e. $d+1$) rank.  
We focus on the most common case in which one of the covariates, say zero, has a special status, and we want to test the null hypothesis that the corresponding coefficient $\gamma_0$ is equal to $0$.
That is, we want to design an e-variable for testing the alternative hypothesis $\cH_1$ vs. the null $\cP$, where $\cH_1$ and $\cP$ are respectively given by: 
\begin{equation}
    \cH_1 = \{G_{\sigma,\vec\gamma} : \vec{\gamma} \in \reals^{d+1},\gamma_0 \neq 0, \sigma > 0 \} 
\ \ \ ; \ \ \ 
\cP=  \{G_{\sigma,\vec\gamma} : \vec{\gamma} \in \reals^{d+1}, \gamma_0 = 0, \sigma > 0 \}.
\end{equation}

We will now show that $\cP$ is an exponential family and $\cH_1$ a union of exponential families, so that we can apply Corollary~\ref{cor:fulldim}.
To this end, define, for $\vec{\gamma} = (\gamma_0, \gamma_1, \ldots, \gamma_d)^T \in \mathbb{R}^{d+1}$ and $\sigma^2\in (0,\infty)$, the conditional normal  distributions $G_{\sigma,\vec{\gamma}}$ with corresponding densities
\begin{equation}\label{eq:lmdens}
g_{\sigma, \vec{\gamma}}(Y^n)  := g_{\sigma,\vec{\gamma}}(Y^n \mid \vec{x}^n) = \frac{1}{(2 \pi \sigma^2)^{n/2}} \cdot e^{-
\frac{1}{2 \sigma^2} \sum (Y_i - \vec{\gamma}^T \vec{x}_i)^2
}.
\end{equation}
Here and in the sequel, sums without explicitly denoted ranges are invariably taken to be over $i=1..n$ and we omit the conditional $\vec{x}^n$ from the notation, since they are fixed throughout the following analysis.
Now define the transformed parameters $\lambda := -1 /(2 \sigma^2)$ and $\vb= (\beta_1, \ldots, \beta_d)^T$ with, for $j=1..d$, $\beta_j := \gamma_j/\sigma^2$ and $\theta:= \gamma_0/\sigma^2$ and set
$t_j(Y^n) = \sum Y_i x_{i,j}$.
Rewriting the likelihood (\ref{eq:lmdens}) in terms of this new parameterization and the $t_j$ and denoting the resulting densities by $f^{(\theta)}_{\lambda,\vec{\beta}}$, we see that
\begin{align}\label{eq:lmexpdens}
    f^{(\theta)}_{\lambda,\vec{\beta}}(y^n) = g_{\sigma,\vec{\gamma}}(y^n) = \exp \left(
    \lambda \sum y_i^2 + \sum_{j=1}^d \beta_j t_j(y^n) \right) \cdot h^{(\theta)}_1(y^n) h^{(\theta)}_2(\sigma,\vec{\gamma})
\end{align}
for some function $h_1^{(\theta)}$ not depending on $(\lambda, \vec{\beta})$ and $h_2^{(\theta)}$ not depending on the data $y^n$. 
Let, for $\theta \in \reals$, $\cQ^{(\theta)}$ be the set of distributions $F^{(\theta)}_{\lambda,\vb}$ with densities $f^{(\theta)}_{\lambda,\vb}$.
We see that, for each $\theta \in \reals$, $\cQ^{(\theta)}$ is a 
$(d+1)$-dimensional exponential family with 
sufficient statistic vector 
\begin{equation}\label{eq:suffstat}
    \left(\sum Y_i^2,  t_1(Y^n), \ldots, t_d(Y^n) \right)
\end{equation}
and mean-value parameter space $\meanspace_\alt^{(\theta)} = \meanspace$, where $\meanspace := (0,\infty) 
\times \reals^d$.
The original parameter vector corresponding to $(\lambda, \vec{\beta})$ is $(\sigma^2, \vec{\gamma})$ with  $\sigma^2 := - 1/(2 \lambda)$ and $\vec{\gamma}= (\sigma^2 \theta, \sigma^2 \beta_1, \ldots, \sigma^2 \beta_d)$ and the corresponding mean-value parameter vector is
\begin{equation}\label{eq:mvagain}
{\vm} := \left(n {\sigma}^2 + \sum (\vec{\gamma}^T \vec{x}_i)^2,
\sum x_{i,1} \vec{\gamma}^T \vec{x}_i, \ldots, \sum x_{i,d} \vec{\gamma}^T \vec{x}_i\right)^T.
\end{equation}
In particular, observe that $\cP= \cQ^{(0)}$ and $\cH_1 = \bigcup_{\theta \in \reals \setminus \{0 \}} \cQ^{(\theta)}$.
For expository convenience,  we slightly deviated from our previous notation here by having a canonical parameter space vector of the form $(\lambda, \vb)$ rather than $\vb= (\beta_1, \ldots, \beta_d)$; thus $\vb$ is $d$-dimensional but $\vm$ still represents a full  $(d+1)$-dimensional mean-value parameter. 

Having established that $\cQ^{(\theta)}$ and $\cP$ are, indeed, exponential families, we will now show that Theorem~\ref{thm:main} in the form of Corollary~\ref{cor:fulldim} is applicable to them.
First, we reparameterize the family of densities in~\eqref{eq:lmexpdens} with respect to a specific $\vm\in \meanspace$ as in~\eqref{eq:exp_fam_dens}.
To this end, let $(\lambda^{(\theta)}(\vm),\vb^{(\theta)}(\vm))$ denote the pair $(\lambda, \vb)$ such that the mean of $F_{\lambda, \vb}^{(\theta)}$ equals $\vm$.
We obtain:
\begin{equation}\label{eq:finalpexpdens}
    q^{(\theta)}_{\lambda, \vb; \vm}(y^n) = \frac{1}{Z_q(\lambda,\vb ;\vm)} \cdot \exp \left(\lambda \sum y_i^2 +  \sum_{j=1}^d \beta_j t_j(y^n)\right) f^{(\theta)}_{\lambda^{(\theta)}(\vm),\vb^{(\theta)}(\vm)}(y^n), 
\end{equation}
with 
$Z_q(\lambda,\vb ; \vm)$  the normalizing constant, defined for all $(\lambda,\vb) \in \canspace^{(\theta)}_{q; \vm}$ where 
$$\canspace^{(\theta)}_{q; \vm} = \{(\lambda,\vb) : Z_p(\lambda,\vb;\vm) < \infty\} = (- \infty, - \lambda^{(\theta)}(\vm)) \times \reals^d.$$ 
We see that the family of densities defined by~\eqref{eq:finalpexpdens} coincides with $\cQ^{(\theta)}$.
Furthermore, we define $p_{\lambda, \vb ;\vm} = q^{(0)}_{\lambda,\vb;\vm}$ for all $(\lambda,\vb) \in \canspace_{p;\vm}$, where $\canspace_{p;\vm}:= \canspace^{(0)}_{q;\vm}$, which corresponds to $\cP$.

As precondition to apply Corollary~\ref{cor:fulldim}, we need to verify  that for each choice of $\theta$, we  have  (i) $\meanspace_q^{(\theta)} \subseteq  \meanspace_p$, and (ii)
for each $\vm \in \meanspace_q^{(\theta)}$, we have that $\canspace_{p;\vm} \subseteq \canspace^{(\theta)}_{q;\vm}$. 
We already verified that $\meanspace_q^{(\theta)}$ is the same for all $\theta$, which implies (i).
As to (ii), note that, for $\vm\in \meanspace$, we have $\canspace_{p;\vm} = (- \infty, - \lambda^{(0)}(\vm)) \times \reals^d$ and $\canspace^{(\theta)}_{q;\vm}= (- \infty, - \lambda^{(\theta)}(\vm)) \times \reals^d$.
It thus remains to be verified that $\lambda^{(0)}(\vm)\geq \lambda^{(\theta)}(\vm) $.
To this end, we will denote $(\sigma^{(\theta)}(\vm),\vec{\gamma}^{(\theta)}(\vm))$ for the original parameter vector corresponding to $(\lambda^{(\theta)}(\vm),\vb^{(\theta)}(\vm))$.
Since the distributions corresponding to $(\lambda^{(\theta)}(\vm),\vb^{(\theta)}(\vm))$ and $(\lambda^{(0)}(\vm),\vb^{(0)}(\vm))$ have the same mean $\vm$, equation~\eqref{eq:mvagain} implies that the following system of equations holds
\begin{align}\label{eq:normaleq}
&  
\text{for all $j \in \{1,\ldots, d\}$:}\ 
\sum (\vg^{(0)}(\vm)^T \vec{x}_i) x_{i,j} {=}
\sum (\vg^{(\theta)}(\vm)^T \vec{x}_i) x_{i,j} \nonumber \\
&    n\sigma^{(0)}(\vm)^2 = n\sigma^{(\theta)}(\vm)^2 + \sum (\vg^{(\theta)}(\vm)^T \vec{x}_i)^2 - \sum (\vg^{(0)}(\vm)^T \vec{x}_i)^2,
\end{align}
which may be seen as versions of the standard {\em normal equations\/} in linear regression analysis. 
It can be verified that
\[ 
\sum (\vg^{(\theta)}(\vm)^T \vec{x}_i)^2 - \sum (\vg^{(0)}(\vm)^T \vec{x}_i)^2 = \sum  (\vg^{(\theta)}(\vm)^T \vec{x}_i - \vg^{(0)}(\vm)^T \vec{x}_i)^2,
\]
which together with~\eqref{eq:normaleq} shows that $\sigma^{(0)}(\vm)^2\geq \sigma^{(\theta)}(\vm)^2$, or equivalently, $\lambda^{(0)}(\vm)\geq \lambda^{(\theta)}(\vm)$, which gives the desired inclusion $\canspace_{p;\vm} \subseteq \canspace^{(\theta)}_{q;\vm}$.

Furthermore, in Appendix~\ref{app:covid}, we establish that $\Sigma_p(\vm)- \Sigma_q^{(\theta)}(\vm)$ is positive semidefinite for all $\theta \in \reals$ and $\vm\in \meanspace$.
Corollary~\ref{cor:fulldim} thus gives that a simple e-variable relative to $G_{\sigma,\vec{\gamma}}$ exists for all $\vec{\gamma} \in \reals^d$ with $\gamma_0 \neq 0$, all $\sigma > 0$.
If $\vm\in \meanspace$ denotes the mean of $G_{\sigma,\vec{\gamma}}$, this simple e-variable is given by $g_{\sigma, \vg}(Y^n)/g_{\sigma^{(0)}(\vm), \vg^{(0)}(\vm)}(Y^n)$.
An e-variable against all of $\cH_1$, or a subset thereof, can be constructed by taking a convex mixture over these simple e-variables, as discussed in Appendix~\ref{app:happyreviewer}.

\section{Proof of Theorem~\ref{thm:main}}
\label{sec:simpleH1generalproofs}
To get some intuition first, we note that the distributions $P_{\vb}$ and $Q_{\vb}$ indexed by the $\vb$ in the definition of $f(\vb;\vm^*)$, i.e. \eqref{eq:logZdifference}, are difficult to compare in the sense that they do not necessarily have any properties in common.
In particular, $P_{\vb}$ generally does not achieve $\min_{P \in \nullhyp} D(Q_{\vb} \| P)$, so that $P_{\vb}$ and $Q_{\vb}$ do not have the same mean.
This suggests to replace $f(\vb;\vm^*)$ by a function $g(\vm;\vm^*)$ on the mean-value parameter space and also to re-express $f(\vb;\vm^*) \leq 0$, the condition for being an e-variable, by a condition on $g$ --- and this is what we do in the proof of Theorem~\ref{thm:main}: inside the proof below we establish, using  well-known convex duality properties of exponential families, that this can be done with function and condition, respectively, given by: 
\begin{align}
    \label{eq:calvados}
g(\vm; \vm^*) &=  D(P_{\vm} \| P_{\vm^*}) - D(Q_{\vm} \| Q_{\vm^*}), \\ 
g(\vm; \vm^*) &\leq 0. \label{eq:g_inequality}
\end{align}
This condition on $g$ corresponds to item~\ref{item:KL} in Theorem~\ref{thm:main}.
The key insight for showing the suitability of $g$ is the following well-known convex-duality fact 
about exponential families: for all $\vm, \vm' \in \meanspace_\nul$, all $\vb \in \canspace_{\nul; \vm^*}$, we have: 
\begin{align}
\label{eq:KLduality}
 -\log Z_{\nul}(\vb; \vm')=D(P_{\vm_{\nul}(\vb; \vm')} \| P_{\vm'}) - \vb^T\vm_{\nul}(\vb; \vm')   \leq 
D(P_{\vm} \| P_{\vm'}) -  \vb^T \vm.
\end{align}
This can be derived as follows:
\begin{align*}
    D(P_{\vm_\nul(\vb;\vm')} \|P_{\vm'})&-D(P_\vm \|P_{\vm'}) = \log \frac{Z_\nul (\beta_\nul (\vm;\vm'))}{Z_\nul(\vb;\vm')}+ \vb^T\vm_\nul(\vb;\vm')-\vb_\nul (\vm;\vm')\vm \\
    &=\log \frac{Z_\nul (\beta_\nul (\vm;\vm'))}{Z_\nul(\vb;\vm')} +\vb^T(\vm_\nul(\vb;\vm')-\vm) -  (\vb_\nul (\vm;\vm')-\vb)^T\vm\\
    &= \vb^T(\vm_\nul(\vb;\vm')-\vm) - D(P_{\vm} \| P_{\vm_\nul(\vb;\vm')})\\
    &\leq \vb^T(\vm_\nul(\vb;\vm')-\vm).
\end{align*}
We now prove the chain of implications in the theorem. 

\paragraph{$(1) \Rightarrow (2)$}
Let $\vm,\vm' \in \meanspace_\alt$ and denote $\vm(\alpha):=(1-\alpha)\vm' + \alpha \vm$.
By assumption of convexity, we have that $\vm(\alpha) \in \meanspace_\alt$ for all $\alpha\in [0,1]$. 
Furthermore, define $h(\alpha)=(\vb_\nul(\vm(\alpha);\vm')-\vb_\alt(\vm(\alpha);\vm'))^T(\vm(\alpha)-\vm')$, so that $h(0)=0$ and $h(1)=(\vb_\nul(\vm;\vm')-\vb_\alt(\vm;\vm'))^T(\vm-\vm')$. 
The derivative of $h$ is given by 
\begin{align*}
    \frac{d}{d\alpha} h(\alpha) =& \left(\frac{d}{d\alpha} \vb_\nul(\vm(\alpha);\vm')-\vb_\alt(\vm(\alpha);\vm')\right)^T(\vm(\alpha)-\vm') \\
    &+ (\vb_\nul(\vm(\alpha);\vm')-\vb_\alt(\vm(\alpha);\vm'))^T \frac{d}{d\alpha} \vm(\alpha).
\end{align*}
The chain rule gives
\begin{align*}
    \frac{d}{d\alpha} \vb_\nul(\vm(\alpha);\vm')
    &=  \Sigma^{-1}_\nul(\vm(\alpha))^T (\vm-\vm'),
\end{align*}
where we use~\eqref{eq:betadiff} and~\eqref{eq:sigmadiff} together with the fact that the Jacobian of the gradient of a function equals the transpose of its Hessian.
The derivative of $\vb_\alt(\vm(\alpha);\vm')$ can be found with the same argument, so we see
\begin{align}
    \frac{d}{d\alpha} h(\alpha) =& \left((\Sigma^{-1}_\nul(\vm(\alpha)) -\Sigma^{-1}_\alt(\vm(\alpha)))^T (\vm-\vm')\right)^T (\vm(\alpha)-\vm')\nonumber \\ 
    &+ (\vb_\nul(\vm(\alpha);\vm')-\vb_\alt(\vm(\alpha);\vm'))^T (\vm-\vm')\nonumber\\
    =&\frac1\alpha (\vm(\alpha)-\vm')^T (\Sigma^{-1}_\nul(\vm(\alpha)) -\Sigma^{-1}_\alt(\vm(\alpha))) (\vm(\alpha)-\vm')\nonumber\\
    &+(\vb_\nul(\vm(\alpha);\vm')-\vb_\alt(\vm(\alpha);\vm'))^T (\vm-\vm')\nonumber\\
    =&\frac1\alpha (\vm(\alpha)-\vm')^T (\Sigma^{-1}_\nul(\vm(\alpha)) -\Sigma^{-1}_\alt(\vm(\alpha))) (\vm(\alpha)-\vm')+\frac1\alpha h(\alpha).\label{eq:hderivative}
\end{align}
If $\Sigma_{\nul}(\vm) - \Sigma_{\alt}(\vm)$ is positive semidefinite for all $\vm$, then $\Sigma^{-1}_{\nul}(\vm)- \Sigma^{-1}_{\alt}(\vm)$ is negative semidefinite (as discussed below the statement of Theorem~\ref{thm:main}). 
In this case, the first term in~\eqref{eq:hderivative} is negative and, since $h(0)=0$, the second term is also negative on $[0,1]$.
It follows that $h$ is decreasing when $\Sigma_{\nul}(\vm) - \Sigma_{\alt}(\vm)$ is positive semidefinite, so that $(\vb_\nul(\vm; \vm' )-\vb_\alt(\vm; \vm'))^T(\vm-\vm'))\leq 0$, as was to be shown. 

\paragraph{$(2) \Rightarrow (3)$}
We use a similar argument as was used to prove the previous implication, so let $\vm,\vm' \in \meanspace_\alt$ and denote $\vm(\alpha)=(1-\alpha)\vm' + \alpha \vm$ as before.
Define $h(\alpha):=g(\vm(\alpha);\vm')$.
Using the chain rule of differentiation together with~\eqref{eq:betadiff}, we see that the derivative of $h$ is given by
\begin{align*}
    \frac{d}{d\alpha} h(\alpha)  &= (\vb_\nul(\vm(\alpha);\vm')-\vb_\alt(\vm(\alpha);\vm'))^T(\vm-\vm')\\ 
    &=\frac{1}{\alpha}(\vb_\nul(\vm(\alpha);\vm')-\vb_\alt(\vm(\alpha);\vm'))^T (\vm(\alpha)-\vm').
\end{align*}
If item~(\ref{item:betas}) holds, then we have that $\frac{d}{d\alpha} h(\alpha)\leq 0$.
Since $h(0)=0$ and $h(1)=D(P_{\vm}\|P_{\vm'})-D(Q_{\vm}\|Q_{\vm'})$, we see that item~(\ref{item:betas}) implies that
\[ D(P_{\vm}\|P_{\vm'})-D(Q_{\vm}\|Q_{\vm'}) \leq 0,\]
as was to be shown.

\paragraph{$(3) \Rightarrow (4)$}
Assume that $D(P_{\vm}\|P_{\vm'})-D(Q_{\vm}\|Q_{\vm'}) \leq 0$ for all $\vm,\vm' \in \meanspace_\alt$.
Together with~\eqref{eq:KLduality} this gives, for all $\vm, \vm' \in \meanspace_\alt$, all $\vb \in \canspace_{\nul; \vm'}$:
\begin{align}
D(P_{\vm_{\nul}(\vb; \vm')} \| P_{\vm'}) - \vb^T \vm_{\nul}(\vb; \vm')  \leq 
D(P_{\vm} \| P_{\vm'}) - \vb^T\vm
\leq  D(Q_{\vm} \| Q_{\vm'}) - \vb^T\vm.
\end{align}
Applying this with $\vm = \vm_{\alt}(\vb; \vm')$ and re-arranging gives 
\begin{align}\label{eq:whiskey}
- D(P_{\vm_{\nul}(\vb; \vm')} \| P_{\vm'}) + \vb^T\vm_{\nul}(\vb; \vm')  \geq 
- D(Q_{\vm_{\alt}(\vb; \vm')} \| Q_{\vm'}) + \vb^T\vm_{\alt}(\vb; \vm'),
\end{align}
which, by the equality in key fact (\ref{eq:KLduality}) is equivalent to  $\log Z_{\nul}(\vb ; \vm') \geq \log Z_{\alt}(\vb; \vm')$, which is what we had to prove. 

\paragraph{Remaining Implications}
$(4) \Rightarrow (5)$ now  follows by the equality in \eqref{eq:logZdifference} and the definition of an e-variable.
$(5) \Rightarrow (6)$ follows from proposition~\ref{prop:GRO},
$(6) \Rightarrow (7)$ follows because a global e-variable is automatically also a local one, and $(7)\Rightarrow (8)$ again follows from Proposition~\ref{prop:GRO}. Finally, $(8)\Rightarrow (1)$ has already been established as Proposition~\ref{prop:local}. \qed

\section{Conclusion and Future Work}\label{sec:conclusion}
We have provided a theorem that, under regularity pre-conditions, provides a general sufficient condition under which there exists a simple e-variable for testing a simple alternative versus a composite regular exponential family null. The characterization was given in terms of several equivalent conditions, the most direct being perhaps the condition
`$\Sigma_{\nul}(\vm) - \Sigma_{\alt}(\vm)$ is positive semidefinite for all $\vm \in \meanspace_\alt$'. 
A direct follow-up question is: can we construct GRO or close-to-GRO e-variables, in case either the regularity pre-conditions or the positive definiteness condition do {\em not\/} hold? 
The example of Section~\ref{sec:nef}, and in particular Figure~\ref{fig:local_evar}, indicated that in that case, many things can happen: under some  $\vm \in \meanspace_q$ (green curve),  $q_{\vm}/p_{\vm}$ still gives a global simple e-variable; for other $\vm$ (blue), it gives a local but not global e-variable; for yet other $\vm$ (pink), it does not give an e-variable at all. 

Nevertheless, it turns out that if the pre-regularity conditions hold and  
the `opposite' of the positive semidefinite condition holds, i.e. if $\Sigma_{\nul}(\vm) - \Sigma_{\alt}(\vm)$ is {\em negative\/} semidefinite for all $\vm \in \meanspace_\alt$,
then there is again sufficient structure to analyze the problem. The GRO e-variable will now be based on a mixture of elements of the null, but the specific mixture will depend on the sample size: we now need to look at i.i.d. repetitions of $U$ rather than a single outcome $U$. We will provide such an analysis in future work. 

Another interesting avenue for future work is to extend the analysis to {\em curved\/} exponential families \citep{efron_2022}. 
While we do not have any general results in this direction yet, the analysis by \cite{liang2023stratified} suggests that this may be possible. 
\citeauthor{liang2023stratified} considers a variation  of the Cochran-Mantel- Haenszel test, in which the null hypothesis expresses that the population-weighted {\em average\/} effect size over a given set of strata is equal to, or bounded by, some $\delta$. This can be rephrased in terms of a curved exponential family null,  for which 
\cite{liang2023stratified} shows that a local e-variable exists by considering the second derivative of the function 
$f(\vb;\vm^*)$ as in (\ref{eq:logZdifference}), just like in the present paper but with $\vb$ representing a particular suitable parameterization rather than the canonical parameterization of an exponential family.  The local e-variable is then shown to be a global e-variable by a technique different from  the construction of $\cQ$ we use here. Still, the overall derivation is sufficiently similar to suggest that it can be unified with the reasoning underlying Theorem~\ref{thm:main}. Finally, the analysis of the linear model in Section~\ref{sec:regression} suggests that the results may perhaps be extended to say something about existence of {\em generalized\/} linear models without assuming a {\em model-X\/} condition \citep{GrunwaldHL22} --- a situation about which currently next to nothing is known.

\DeclareRobustCommand{\VANDER}[3]{#3}
\bibliography{references,SAVIreferences}

\begin{thebibliography}{43}
\providecommand{\natexlab}[1]{#1}
\providecommand{\url}[1]{\texttt{#1}}
\expandafter\ifx\csname urlstyle\endcsname\relax
  \providecommand{\doi}[1]{doi: #1}\else
  \providecommand{\doi}{doi: \begingroup \urlstyle{rm}\Url}\fi

\bibitem[Agrawal(2018)]{Agrawal18}
Akshay Agrawal.
\newblock Lecture notes on {L}oewner order, 2018.
\newblock URL \url{https://www.akshayagrawal.com/lecture-notes/html/loewner-order.html}.

\bibitem[Anderson and Darling(1954)]{anderson1954test}
Theodore~W Anderson and Donald~A Darling.
\newblock A test of goodness of fit.
\newblock \emph{Journal of the American statistical association}, 49\penalty0 (268):\penalty0 765--769, 1954.

\bibitem[Awad et~al.(2022)Awad, Bar-Lev, and Makov]{awad2022new}
Yaser Awad, Shaul~K Bar-Lev, and Udi Makov.
\newblock A new class of counting distributions embedded in the {L}ee--{C}arter model for mortality projections: A {B}ayesian approach.
\newblock \emph{Risks}, 10\penalty0 (6):\penalty0 111, 2022.

\bibitem[Bar-Lev(2020)]{bar2020independent}
Shaul~K Bar-Lev.
\newblock Independent, though identical results: the class of {T}weedie on power variance functions and the class of {B}ar-{L}ev and {E}nis on reproducible natural exponential families.
\newblock \emph{Int. J. Stat. Probab}, 9\penalty0 (1):\penalty0 30--35, 2020.

\bibitem[Bar-Lev and Ridder(2021)]{BarLevR21}
Shaul~K Bar-Lev and Ad~Ridder.
\newblock New exponential dispersion models for count data -- the {ABM} and {LM} classes.
\newblock \emph{ESAIM: Probability and Statistics}, 25:\penalty0 31--52, 2021.

\bibitem[Bar-Lev and Ridder(2023)]{bar2023exponential}
Shaul~K Bar-Lev and Ad~Ridder.
\newblock Exponential dispersion models for overdispersed zero-inflated count data.
\newblock \emph{Communications in Statistics-Simulation and Computation}, 52\penalty0 (7):\penalty0 3286--3304, 2023.

\bibitem[Bar-Lev et~al.(2024)Bar-Lev, Letac, and Ridder]{BarlevLR24}
Shaul~K Bar-Lev, G{\'e}rard Letac, and Ad~Ridder.
\newblock A delineation of new classes of exponential dispersion models supported on the set of nonnegative integers.
\newblock \emph{Annals of the Institute of Statistical Mathematics}, 2024.
\newblock URL \url{https://doi.org/10.1007/s10463-024-00903-y}.

\bibitem[Barndorff-Nielsen(1978)]{BarndorffNielsen78}
O.E. Barndorff-Nielsen.
\newblock \emph{Information and Exponential Families in Statistical Theory}.
\newblock Wiley, Chichester, UK, 1978.

\bibitem[Barron et~al.(1998)Barron, Rissanen, and Yu]{BarronRY98}
A.~Barron, J.~Rissanen, and B.~Yu.
\newblock The minimum description length principle in coding and modeling.
\newblock \emph{IEEE transactions on information theory}, 44\penalty0 (6):\penalty0 2743--2760, 1998.
\newblock Special Commemorative Issue: Information Theory: 1948-1998.

\bibitem[Barron and Sheu(1991)]{BarronS91}
A.R. Barron and C.~Sheu.
\newblock Approximation of density functions by sequences of exponential families.
\newblock \emph{Annals of Statistics}, 19\penalty0 (3):\penalty0 1347--1369, 1991.

\bibitem[Brinda(2018)]{Brinda18}
William~David Brinda.
\newblock \emph{Adaptive Estimation with Gaussian Radial Basis Mixtures}.
\newblock PhD thesis, Yale University, 2018.

\bibitem[Brown(1986)]{brown1986fundamentals}
Lawrence~D. Brown.
\newblock Fundamentals of statistical exponential families with applications in statistical decision theory.
\newblock \emph{Lecture Notes-Monograph Series}, 9:\penalty0 i--279, 1986.
\newblock ISSN 07492170.
\newblock URL \url{http://www.jstor.org/stable/4355554}.

\bibitem[Clarke and Barron(1990)]{ClarkeB90}
B.S. Clarke and A.R. Barron.
\newblock Information-theoretic asymptotics of {B}ayes methods.
\newblock \emph{IEEE Transactions on Information Theory}, 36\penalty0 (3):\penalty0 453--471, 1990.
\newblock \doi{10.1109/18.54897}.

\bibitem[Efron(2022)]{efron_2022}
Bradley Efron.
\newblock \emph{Exponential Families in Theory and Practice}.
\newblock Institute of Mathematical Statistics Textbooks. Cambridge University Press, 2022.
\newblock \doi{10.1017/9781108773157}.

\bibitem[Gr{\"u}nwald(2007)]{grunwald2007minimum}
Peter Gr{\"u}nwald.
\newblock \emph{The minimum description length principle}.
\newblock MIT press, 2007.

\bibitem[Gr{\"u}nwald(2024)]{Grunwald24}
Peter Gr{\"u}nwald.
\newblock Beyond {N}eyman-{P}earson: e-values enable hypothesis testing with a data-driven alpha.
\newblock \emph{Proceedings National Academy of Sciences of the USA (PNAS)}, 121\penalty0 (39), 2024.

\bibitem[Gr{\"u}nwald et~al.(2024)Gr{\"u}nwald, Heide, and Koolen]{GrunwaldHK23}
Peter Gr{\"u}nwald, Rianne~\VANDER{Heide}{De}{de} Heide, and Wouter Koolen.
\newblock Safe testing.
\newblock \emph{Journal of the Royal Statistical Society, Series B}, 2024.
\newblock with discussion.

\bibitem[Grünwald et~al.(2023)Grünwald, Henzi, and Lardy]{GrunwaldHL22}
Peter Grünwald, Alexander Henzi, and Tyron Lardy.
\newblock Anytime-valid tests of conditional independence under model-{X}.
\newblock \emph{Journal of the American Statistical Association}, 0\penalty0 (0):\penalty0 1--12, 2023.

\bibitem[Hao and Grünwald(2024)]{hao2024expfamgeneral}
Yunda Hao and Peter Grünwald.
\newblock E-values for exponential families: the general case, 2024.
\newblock URL \url{https://arxiv.org/abs/2409.11134}.
\newblock arXiv preprint arXiv:2409.11134.

\bibitem[Hao et~al.(2024)Hao, Gr{\"u}nwald, Lardy, Long, and Adams]{HaoGLLA23}
Yunda Hao, Peter Gr{\"u}nwald, Tyron Lardy, Long Long, and Reuben Adams.
\newblock E-values for k-sample tests with exponential families.
\newblock \emph{Sankhya A}, 2024.

\bibitem[J{{\o{}}rgensen}(1997)]{Jorgensen97}
Bent J{{\o{}}rgensen}.
\newblock \emph{The Theory of Exponential Dispersion Models}, volume~76 of \emph{Monographs on Statistics and Probability}.
\newblock Chapman and Hall, London, 1997.

\bibitem[Koolen and Gr{\"u}nwald(2021)]{KoolenG21}
W.~Koolen and P.~Gr{\"u}nwald.
\newblock Log-optimal anytime-valid e-values.
\newblock \emph{International Journal of Approximate Reasoning}, 2021.
\newblock Festschrift for G. Shafer's 75th Birthday.

\bibitem[Lardy(2021)]{Lardy21}
Tyron Lardy.
\newblock E-values for hypothesis testing with covariates, 2021.
\newblock Master's {T}hesis, {L}eiden {U}niversity.

\bibitem[Lardy et~al.(2024)Lardy, Gr\"unwald, and Harremo\"es]{LardyHG24}
Tyron Lardy, Peter Gr\"unwald, and Peter Harremo\"es.
\newblock Reverse information projections and optimal e-statistics.
\newblock \emph{IEEE Transactions on Information Theory}, 70\penalty0 (11):\penalty0 7616--7631, 2024.
\newblock \doi{https://doi,org/10.1109/TIT.2024.3444458}.

\bibitem[Larsson et~al.(2025)Larsson, Ramdas, and Ruf]{larsson2024numeraire}
Martin Larsson, Aaditya Ramdas, and Johannes Ruf.
\newblock The num{\'e}raire e-variable and reverse information projection.
\newblock \emph{Annals of Statistics}, 2025.

\bibitem[Li and Barron(2000)]{LiB00}
J.Q. Li and A.R. Barron.
\newblock Mixture density estimation.
\newblock In \emph{Advances in Neural Information Processing Systems}, volume~12, pages 279--285, 2000.

\bibitem[Li(1999)]{li1999estimation}
Qiang~Jonathan Li.
\newblock \emph{Estimation of mixture models}.
\newblock Yale University, 1999.

\bibitem[Liang(2023)]{liang2023stratified}
Haoyi Liang.
\newblock Stratified safe sequential testing for mean effect, 2023.
\newblock Master's Thesis, University of Amsterdam.

\bibitem[Lilliefors(1967)]{lilliefors1967kolmogorov}
Hubert~W Lilliefors.
\newblock On the {K}olmogorov-{S}mirnov test for normality with mean and variance unknown.
\newblock \emph{Journal of the American statistical Association}, 62\penalty0 (318):\penalty0 399--402, 1967.

\bibitem[Lindon et~al.(2024)Lindon, Ham, Tingley, and Bojinov]{lindon2024anytime}
Michael Lindon, Dae~Woong Ham, Martin Tingley, and Iavor Bojinov.
\newblock \emph{Anytime-valid Inference in Linear Models and Regression-adjusted Inference}.
\newblock Harvard Business School, 2024.

\bibitem[Ly et~al.(2025)Ly, Boehm, Gr\"unwald, Ramdas, and van Ravenzwaaij]{Ly25}
Alexander Ly, Udo Boehm, Peter Gr\"unwald, Aaditya Ramdas, and Don van Ravenzwaaij.
\newblock A tutorial on safe anytime-valid inference: Practical maximally flexible sampling designs for experiments based on e-values, 2025.
\newblock PsyArXiv Preprint.

\bibitem[Morris(1982)]{morris1982natural}
Carl~N Morris.
\newblock Natural exponential families with quadratic variance functions.
\newblock \emph{The Annals of Statistics}, pages 65--80, 1982.

\bibitem[P{\'e}rez-Ortiz et~al.(2024)P{\'e}rez-Ortiz, Lardy, {D}e Heide, and Gr{\"u}nwald]{perez2024estatistics}
Muriel~Felipe P{\'e}rez-Ortiz, Tyron Lardy, Rianne {D}e Heide, and Peter Gr{\"u}nwald.
\newblock E-statistics, group invariance and anytime valid testing.
\newblock \emph{Annals of Statistics}, 2024.

\bibitem[Ramdas and Wang(2025)]{ramdas2025hypothesistestingevalues}
Aaditya Ramdas and Ruodu Wang.
\newblock Hypothesis testing with e-values, 2025.
\newblock URL \url{https://arxiv.org/abs/2410.23614}.

\bibitem[Ramdas et~al.(2023)Ramdas, Grünwald, Vovk, and Shafer]{ramdas2023savi}
Aaditya Ramdas, Peter Grünwald, Vladimir Vovk, and Glenn Shafer.
\newblock Game-theoretic statistics and safe anytime-valid inference.
\newblock \emph{Statistical Science}, 2023.

\bibitem[Spector et~al.(2023)Spector, Candès, and Lei]{SpectorCL23}
Asher Spector, Emmanuel Candès, and Lihua Lei.
\newblock Discussion of “a note on universal inference” by {T}se and {D}avison.
\newblock \emph{Stat}, 12\penalty0 (1):\penalty0 e570, 2023.
\newblock \doi{https://doi.org/10.1002/sta4.570}.
\newblock URL \url{https://onlinelibrary.wiley.com/doi/abs/10.1002/sta4.570}.

\bibitem[Stephens(1974)]{stephens1974edf}
Michael~A Stephens.
\newblock {EDF} statistics for goodness of fit and some comparisons.
\newblock \emph{Journal of the American statistical Association}, 69\penalty0 (347):\penalty0 730--737, 1974.

\bibitem[Takeuchi and Barron(1997)]{TakeuchiB97}
J.~Takeuchi and A.~Barron.
\newblock Asymptotically minimax regret for exponential families.
\newblock In \emph{Proceedings SITA '97}, pages 665--668, 1997.

\bibitem[Takeuchi and Barron(2024)]{TakeuchiB24}
Jun-Ichi Takeuchi and Andrew Barron.
\newblock Asymptotic minimax regret for {B}ayes mixtures.
\newblock \emph{arXiv preprint arxiv:2406.17929}, 2024.

\bibitem[Turner et~al.(2024)Turner, Ly, and Gr{\"u}nwald]{TurnerLG24}
Rosanne Turner, Alexander Ly, and Peter Gr{\"u}nwald.
\newblock Generic e-variables for exact sequential k-sample tests that allow for optional stopping.
\newblock \emph{Statistical Planning and Inference}, 230, 2024.

\bibitem[\VANDER{Jong}{De}{de}~Jong(2021)]{DeJong21}
Martijn \VANDER{Jong}{De}{de}~Jong.
\newblock Tests of significance for linear regression using {E}-values, 2021.
\newblock Master's {T}hesis, {L}eiden {U}niversity.

\bibitem[Wang et~al.(2025)Wang, Wang, and Ziegel]{wang2023ebacktesting}
Qiuqi Wang, Ruodu Wang, and Johanna Ziegel.
\newblock E-backtesting.
\newblock \emph{Management Science}, 2025.
\newblock To Appear.

\bibitem[Zhang et~al.(2023)Zhang, Ramdas, and Wang]{zhang2023existence}
Zhenyuan Zhang, Aaditya Ramdas, and Ruodu Wang.
\newblock On the existence of powerful p-values and e-values for composite hypotheses, 2023.
\newblock arXiv preprint arXiv:2304.16539.

\end{thebibliography}

\appendix
\section{Illustrating the Use of A Simple E-Variable}
\label{app:happyreviewer}
In this appendix, we aim to illuminate two things: (1) how simple e-variables can be used to construct e-variables for composite alternatives, and (2) how simple e-variables can be used to construct anytime-valid sequential tests.
To illustrate these points, we focus on the setting of Section~\ref{sec:moreksample} with $k=2$: two-sample tests with Bernoulli data.
This setting was previously studied in detail by~\citet{TurnerLG24}; we refer to their work for more details.

\subsection{Dealing with Composite Alternative}\label{app:composite}
\begin{figure}[ht!]
    \centering
    \includegraphics[width=0.65\linewidth]{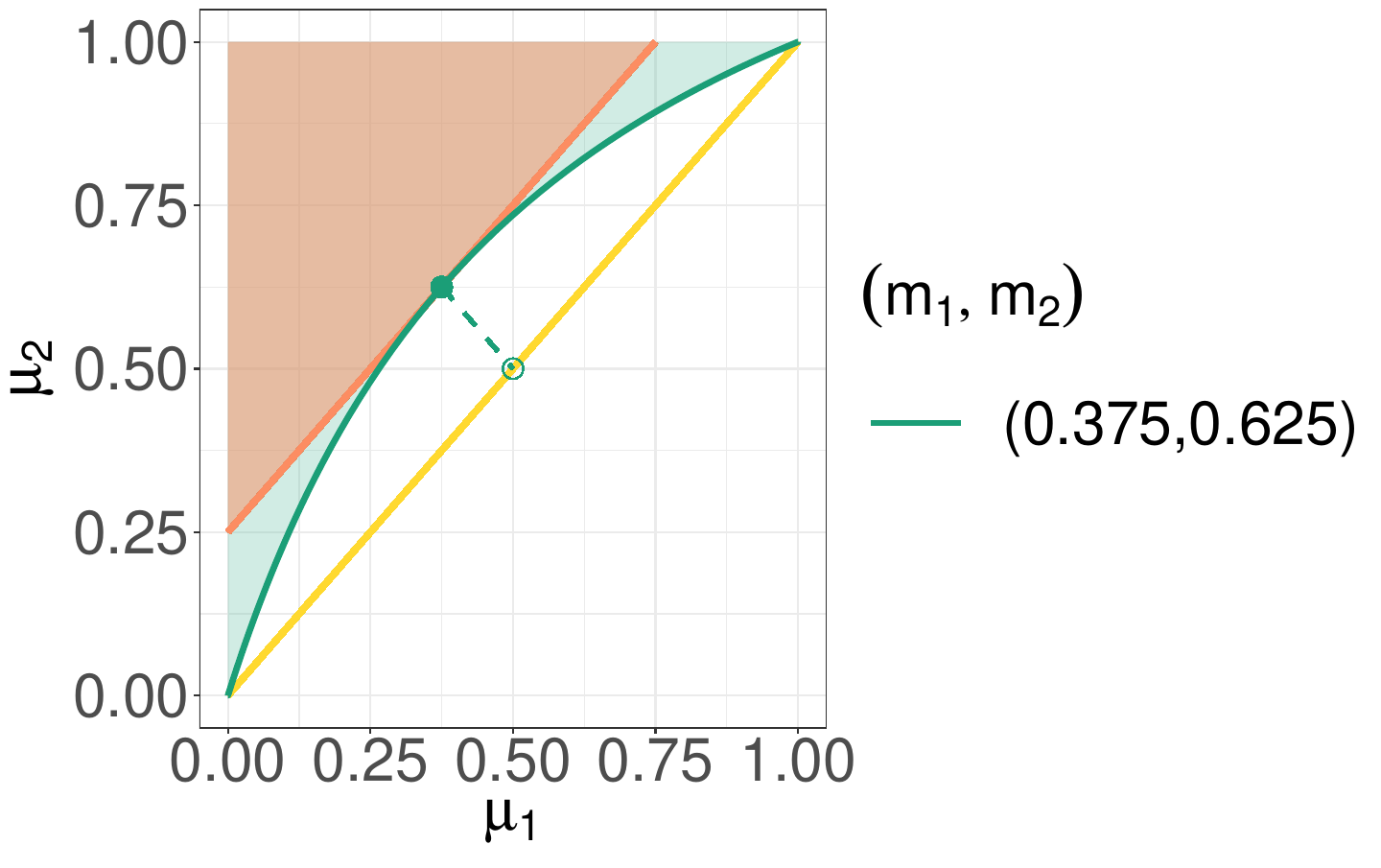}
    \caption{The coordinate grid represents the parameters of the full $2$-sample Bernoulli family. 
    The straight yellow line shows the parameter space of $\nullhyp$.
    The green point represents a single alternative point $(m_1,m_2)=(0.375,0.625)$.
    The shaded red area represents the effect size $\mu_2-\mu_1 > m_2-m_1=0.25$ and the shaded green and red areas together represent $\log( \frac{\mu_2}{1-\mu_2} \frac{1-\mu_1}{\mu_1})>\log( \frac{m_2}{1-m_2} \frac{1-m_1}{m_1})$.
    The curved green line shows the parameters of the distributions in $\althyp$ (which coincides with the boundary of the effect size region).
    Finally, the dashed line shows the projection of $(m_1,m_2)$ onto the parameter space of $\nullhyp$.}
    \label{fig:bernoullieffsize}
\end{figure}

In the two-sample problem, we observe data $(Y_1,Y_2)$, where $Y_1$ and $Y_2$ are independently Bernoulli distributed with means $\mu_1,\mu_2\in (0,1)$, respectively.
The null hypothesis is given by $\cP: \mu_1=\mu_2$ and a common alternative is $\cH_1: d(\mu_1,\mu_2)> \delta$, where
$d(\mu_1,\mu_2)$ is some notion of effect size;
popular choices are either $d(\mu_1,\mu_2)=\mu_2-\mu_1$ or $d(\mu_1,\mu_2)=\log( \frac{\mu_2}{1-\mu_2} \frac{1-\mu_1}{\mu_1})$.
Although $\cP$ itself is composite, the results in Section~\ref{sec:moreksample} can be used to conclude that, for each $Q_{m_1,m_2}\in \cH_1$ with mean $(m_1,m_2)\in (0,1)^2$, the statistic $S_{m_1,m_2}(Y_1,Y_2)$ is a simple e-variable, where
\begin{equation}\label{eq:simplealt}
 S_{m_1,m_2}(Y_1,Y_2)=\frac{q_{m_1,m_2}(Y_1,Y_2)}{p_{m^*}(Y_1,Y_2)}= \frac{m_1^{Y_1}(1-m_1)^{1-Y_1} m_2^{Y_1}(1-m_2)^{1-Y_2}}{m^{* Y_1+Y_2}(1-m^*)^{2-(Y_1+Y_2)} }.
 \end{equation}
Here, $P_{m^*}$ denotes the distribution in $\cP$ under which $Y_1+Y_2$ has mean $m^*=m_1+m_2$ (i.e. $Y_1,Y_2$ are i.i.d. Bernoulli $m^*/2$).
This problem is visualized in Figure~\ref{fig:bernoullieffsize}.
Note in particular that the constructed family $\cQ$ does not coincide with (a subset of) $\cH_1$ if we define $d(\mu_1,\mu_2)=\mu_2-\mu_1$, while it does if we define $d(\mu_1,\mu_2)=\log( \frac{\mu_2}{1-\mu_2} \frac{1-\mu_1}{\mu_1})$.
Nevertheless, we can use a standard construction, {\em Robbins'  method of mixture}, to build an e-variable to test the null $\cP$ against $\cH_1$ regardless of the definition of $d(\mu_1,\mu_2)$. While the resulting e-variable will not be very powerful, either in terms of standard power or in terms of growth rate (GRO as in the main text), as long as we restrict ourselves  to samples of size 1 (a single pair $(Y_1,Y_2)$, it will become powerful and close to growth rate optimal once we extend the construction to larger samples.

We thus start by considering a single pair. Note that the e-variable $S_{m_1,m_2}(Y_1,Y_2)$ can be thought of as a measure of evidence against $\cP$ with respect to $Q_{m_1,m_2}$.
To turn this into evidence with respect to (a subset of) $\cH_1$, we can pick any prior $W$ on $\meanspace_\alt = \{(m_1,m_2): m_2-m_1>\delta \}$, and construct an e-variable as follows
\begin{equation}\label{eq:priorevar}
    S_W(Y_1,Y_2)=\int_{\meanspace_\alt} S_{m_1,m_2}(Y_1,Y_2) \, \mathrm{d}W(m_1,m_2).
\end{equation}
The fact that $S_W$ is an e-variable is easily verified by
\[\mathbb{E}_P[S_W] = \mathbb{E}_P\left[\int_{\meanspace_\alt} S_{m_1,m_2}(Y_1,Y_2) \, \mathrm{d}W(m_1,m_2)\right] = \int_{\meanspace_\alt}  \mathbb{E}_P[S_{m_1,m_2}(Y_1,Y_2) ]\, \mathrm{d}W(m_1,m_2) \leq 1,\]
where we use the Fubini-Tonelli theorem together with the fact that each $S_{m_1,m_2}(Y_1,Y_2)$ is an e-variable.
The e-variable $S_W$ is a measure of evidence against $\cP$ with respect to (a subset of) $\cH_1$ weighted by the prior, showing how e-variables for simple alternatives can be used to construct e-variables for composite alternatives.

\subsection{Dealing with Data Sequences and a Composite Alternative}\label{app:sequential}
Above, the data was given by a single pair $(Y_1,Y_2)$.
In this case, it follows from the discussion in Section~\ref{sec:evars} that the e-variable $S_W(Y_1,Y_2)$ (as in~\eqref{eq:priorevar}) can be used to construct a hypothesis test for $\cP$ by defining $\phi(Y_1,Y_2)= \mathbf{1}\left\{S_W(Y_1,Y_2)\geq \frac{1}{\alpha} \right\}$.
The latter is a valid test at significance level $\alpha$, meaning that $P(\phi(Y_1,Y_2)=1)\leq \alpha$ for all $P\in \cP$.
However, it is defined in a static setting, while the main application of e-variables is in anytime-valid settings.

Suppose then that we are in a more dynamic setting where we do not observe a single pair $(Y_1,Y_2)$, but a stream of i.i.d. data points $(Y_{1,1},Y_{1,2}),(Y_{2,1},Y_{2,2}),\dots$. 
It can easily be seen that, for any fixed $Q_{m_1,m_2}$ in $\cH_1$, it holds that $S_{m_1,m_2}(Y_{n,1},Y_{n,2})$ as in (\ref{eq:simplealt}) is an e-variable for each $n\in \mathbb{N}$.
This e-variable measures the evidence against $\cP$ gathered in the $n$th round.
As a measure of the accumulated evidence combined over all the data points up to time $n$, we may consider the product $S^{(n)}_{m_1,m_2} =\prod_{i=1}^n S_{m_1,m_2}(Y_{i,1}, Y_{i,2})$.
By independence of the data points, it is straightforward to see that $S^{(n)}_{m_1,m_2}$ is an e-variable for each $n$.
It turns out that a stronger statement is also true~\cite[see e.g.][]{ramdas2023savi, GrunwaldHK23}:
setting $S_n := S^{(n)}_{m_1,m_2}$, we have {\em Ville's inequality\/}
\begin{equation}\label{eq:ville} P\left( \exists n\in \mathbb{N}: S_n \geq \frac{1}{\alpha}\right)\leq \alpha \quad \text{for all} \quad P\in\cP .\end{equation}
It follows that the sequence of tests $(\phi_n)_{n\in \mathbb{N}}$ defined by $\phi_n(Y_{1,1},Y_{1,2},\dots,Y_{n,1},Y_{n,2})= \mathbf{1}\left\{S_n\geq \frac{1}{\alpha} \right\}$ has a type-I error probability that is uniformly bounded by $\alpha$ over time.
That is, $P(\exists n\in \mathbb{N}: \phi_n(Y_{1,1},Y_{1,2},\dots,Y_{n,1},Y_{n,2})=1)\leq \alpha$ for all $P\in \cP$.
This is what is referred to as an anytime-valid test~\citep{ramdas2023savi}.
This shows the usefulness of e-variables, and hence our results, in a sequential setting, for a simple alternative $Q_{m_1,m_2}$. For a composite alternative $\cH_1$, we may apply the construction of (\ref{eq:priorevar}) and set
\begin{equation}\label{eq:priorevarb}
    S^{(n)}_W =\int_{\meanspace_\alt} S^{(n)}_{m_1,m_2} \, \mathrm{d}W(m_1,m_2).
\end{equation}
This is again easily checked to be an e-variable for each $n$, and it turns out that Ville's inequality (\ref{eq:ville}) holds again with $S_n := S^{(n)}_W $, so we can use it to define an anytime-valid test of $\cP$ against alternative $\cH_1$. The use of a prior $W$ on the alternative now ensures that such a test will have power against all of $\cH_1$, since for every element $Q_{m_1,m_2}$ of $\cH_1$ we have the following: as the sample size increases, if data are sampled from $Q_{m_1,m_2}$ (which is hence the `true' alternative instance), $S^{(n)}_W$ starts to behave more and more similarly to the oracle test $S^{(n)}_{m_1,m_2}$ that was designed specifically for this `true' (in reality unknown) alternative. 
This is easiest to see if we take $W$ under which $m_1$ and $m_2$ are independently beta-distributed. In that case \citep{TurnerLG24} we we can write $S^{(n)}_W = \prod_{i=1}^n S_{\breve{m}_{1|i},\breve{m}_{2|i}}(Y_{i,1}, Y_{i,2})$ with $\breve{m}_{1|i},\breve{m}_{2|i}$ the posterior mean estimates, after observing $(Y_{1,1}, Y_{1,2}, \ldots, Y_{i-1,1}, Y_{i-1,2})$, of $m_1$ and $m_2$ respectively. Thus, just like in Bayesian inference, in the case the alternative is true, the use of a prior enables us to {\em learn\/} the unknown alternative instance (parameters) $(m_1,m_2)$, and for large $n$, the resulting e-variable will be nearly as large as  $S^{(n)}_{m_1,m_2}$ based on the unknown true instances (the precise relation between $S^{(n)}_W$ and $S^{(n)}_{m_1,m_2}$ is quantified by \cite{hao2024expfamgeneral}). Still, it should be noted that the sequence of e-variables is {\em valid}, and hence the anytime-valid Type-I error guarantee (\ref{eq:ville}) holds, irrespective of the choice of prior $W$. \cite{TurnerLG24} illustrate this particular case, Bernoulli two-sample tests, using a real-world example: the infamous SWEPIS clinical trial, that was stopped early for harm; because the original sampling plan had not been followed, the resulting evidence (a p-value) was statistically invalid.  If the trialists had used e-processes such as $S_W^{(n)}$ instead of $p$-values, they could have stopped even earlier, perhaps thereby even  saving a life, and would have nevertheless have ended with a valid statistical conclusion. 

The general methodology of starting with a set of e-variables, one for each sub-alternative in $\cH_1$, and then constructing a new e-variable by averaging over a prior over all resultant e-variables is known as {\em Robbins' method of mixtures}. It can be applied in wide generality \citep{ramdas2023savi,GrunwaldHK23}, invariably leading to an anytime-valid test via (\ref{eq:ville}). 

\section{Details for Section~\ref{sec:regression}}
\label{app:covid}
We need to establish that $\Sigma_p(\vm)- \Sigma_q^{(\theta)}(\vm)= \Sigma_q^{(0)}(\vm)- \Sigma_q^{(\theta)}(\vm)$ is positive semidefinite for all $\vm \in \reals^d$. 

Thus, take any $\vm^*\in \reals^d$.
For the sake of brevity, we will denote $\lambda^*= \lambda^{(0)}(\vm), \vb^*=\vb^{(0)}(\vm), \lambda^{\circ}= \lambda^{(\theta)}(\vm)$ and $\vb^{\circ}= \vb^{(\theta)}(\vm)$, so that $F^{(0)}_{\lambda^*,\vb^*}$ and $F^{(\theta)}_{\lambda^\circ, \vb^\circ}$ are the distributions in $\cP$ and $\althyp^{(\theta)}$, respectively, with mean $\vm$.
By (\ref{eq:mvagain}), we have that $\lambda^{\circ}, \vb^{\circ}$ and $\hilde\lambda,\hilde{\vb}$ that are related to each other via the normal equations (\ref{eq:normaleq}). 
Furthermore, based on the sufficient statistics (\ref{eq:suffstat}), we can write, for $\theta' \in \{0,\theta \}$, that 
$$
\Sigma_q^{(\theta')}(\vm^*) = \begin{pmatrix} A^{(\theta')} & B^{(\theta')} \\
    (B^{(\theta')})^T  & C^{(\theta')}
\end{pmatrix}
$$
where $A^{(\theta)}$ is the variance of $\sum Y_i^2$ according to
distribution $F^{(\theta)}_{\lambda^{\circ},\vb^{\circ}}$ and $C^{(\theta)}$ is the $d \times d$ covariance matrix of the $t_j(Y^n)$ according to this distribution
and 
$$B^{(\theta)} = \left(\textsc{cov} \left(\sum Y_i^2, t_1(Y^n)\right), \ldots, \textsc{cov}\left(\sum Y_i^2, t_d(Y^n) \right) \right)$$ 
where the covariances are again under this distribution.
Similarly, $A^{(0)}$ is the variance of $\sum Y_i^2$ according to
distribution $F^{(0)}_{\hilde\lambda,\hilde\vb}$ 
and $B^{(0)}$, $C^{(0)}$ are defined accordingly.

Positive semidefiniteness of $\Sigma_q^{(0)}(\vm^*) - \Sigma_q^{(\theta)}(\vm^*)$ is easily seen to be implied\footnote{For an explicit derivation see {\tt https://math.stackexchange.com/questions/2280671/} \\ {\tt definiteness-of-a-general-partitioned-matrix-mathbf-m-left-beginmatrix-bf}.}
if we can show that $C^{(0)} - C^{(\theta)}$ is positive definite and that 
\begin{equation}\label{eq:final}
(A^{(0)}- A^{(\theta)}) - (B^{(0)}-B^{(\theta)})^T (C^{(0)} - C^{(\theta)})^{-1} (B^{(0)} - B^{(\theta)}) \geq 0.
\end{equation}
To show that $C^{(0)} - C^{(\theta)}$ is positive definite, note that $C^{(\theta)}$
%
is simply the standard covariance matrix in linear regression scaled by $1/\sigma^{\circ 2}$, i.e.
$C^{(\theta)} = \sigma^{\circ 2} \sum {\bf x}_i {\bf x}_i^T$ which by the maximal rank assumption is positive definite. Similarly
$C^{(0)}= \sigma^{* 2} \sum {\bf x}_i {\bf x}_i^T$
so that, since by assumption $\theta \neq 0$ and using the normal equations (\ref{eq:normaleq}), we have that  
$C^{(0)} - C^{(\theta)}=  c \sum {\bf x}_i {\bf x}_i^T$ 
for $c = \sigma^{ *2 } - \sigma^{\circ 2} > 0$ is also positive definite.

It only remains to show (\ref{eq:final}).
For the sake of brevity, we denote $\nu_i^*= \vg^{* T} \vec{x}_i$ and $\nu_i^\circ = \vg^{\circ T} \vec{x}_i$, where $\vg^*$ and $\vg^\circ$ are the original parameters corresponding to $(\lambda^*,\vb^*)$ and $(\lambda^\circ,\vb^\circ)$, respectively.
As established by considering the moments of the normal distribution,
%
we have that $A^{(\theta)}=  2\sigma^{\circ 2} \left(2 (\sum \nu_i^{\circ 2}) + n \sigma^{\circ 2}\right)$ and 
$A^{(0)}=  2\sigma^{* 2} \left(2 (\sum \nu^{* 2}_i) +  n \sigma^{* 2}\right) $.
Similarly, we find 
$B^{(\theta)}_j = 2 \sigma^{\circ 2} \left(\sum \nu_i^{\circ} x_{i,j} \right)$ 
and similarly $B^{(0)}_j = 2 \sigma^{* 2} \left(\sum \nu^*_i x_{i,j} \right)$.
By the normal equations (\ref{eq:normaleq}) we find that 
$B^{(0)}_j - B^{(\theta)}_j = 2 (\sigma^{*2} - \sigma^{\circ 2}) \sum \hilde{\nu}_i x_{i,j}$.
After some matrix multiplications (where we may use the cyclic property of the trace of a matrix product) we get that (\ref{eq:final}) is equivalent to 
$$
(A^{(0)}
- A^{(\theta)}) - 4 (\sigma^{*2} - \sigma^{ \circ 2 }) 
\sum \nu^{* 2}_i \geq 0.
$$
But this is easily verified: it is equivalent to 
$$
2 \sigma^{*2} \left(  2\left( 
\sum\nu_i^{* 2} \right) + n \sigma^{*2} - 2\left( 
\sum\nu_i^{* 2} \right)  \right) - 
2 \sigma^{\circ 2} \left(  2\left( 
\sum\nu_i^{\circ 2} \right) + n \sigma^{\circ 2} - 2\left( 
\sum\nu_i^{* 2} \right)  \right) \geq 0
$$
which in turn is equivalent to 
$$
2 n \sigma^{* 4} - 2 n \sigma^{\circ 4}  +4  (\sum\nu_i^{* 2}- \sum\nu_i^{\circ 2}) \sigma^{\circ 2} \geq 0 
$$
which by the normal equations is equivalent to 
$$
 \sigma^{* 4} -  \sigma^{\circ 4}  + 2 (\sigma^{*2} - \sigma^{\circ 2} ) \sigma^{\circ 2} \geq 0 
$$
but this must be the case since by the normal equations, $\sigma^{*2} > \sigma^{\circ 2}$.

\end{document}